# Beyond the Mean-flow: A Spectral-Dynamic Approach to Unraveling the Physics of Droplet Capture in Fog Harvesting Meshes

Pradyumna Das[1], Ranjan Ganguly[2], and Ashoke De[3*]

[1]*Department of Sustainable Energy Engineering, IIT Kanpur, Kanpur – 208016, India*

[2]*Department of Power Engineering, Jadavpur University, Kolkata – 700098, India*

[3]*Department of Aerospace Engineering, IIT Kanpur, Kanpur – 208016, India.*

**corresponding author: ashoke@iitk.ac.in*

Fog harvesting efficiency with mesh collectors is governed by complex interactions between droplet inertia and geometry-induced flow structures. Although previous studies have primarily relied on mean-flow metrics, the present work introduces a spectral-dynamic framework to examine an important but often overlooked control on droplet capture. A two-way coupled Eulerian-Lagrangian model is used to simulate droplet-laden flow (2-40 μm) across five representative mesh geometries. The results show that capture efficiency correlates not only with the magnitude of velocity fluctuations, but also with their spectral distribution and persistence. Frequency-domain analysis indicates that mesh geometry redistributes fluctuation energy across pore and obstruction regions, thereby defining a characteristic flow timescale. By comparing this flow timescale with the droplet response time, a dynamic matching parameter, $\Pi = droplet\ response\ time/flow\ time\ scale$, is introduced. The highest capture efficiency occurs when Π is order unity, corresponding to sustained droplet-flow interaction in the near-mesh region. Geometries that generate broadband, moderately amplified spectral content (e.g., triangular mesh) increase droplet residence time and interception probability, whereas geometries with either weak or highly localized fluctuations reduce performance through insufficient forcing or premature bypass. A physics-inspired correlation for capture efficiency is proposed based on this condition. The study therefore provides mechanistic design guidance, rather than a definitive optimum, for geometry optimization in fog-harvesting meshes.

## 1. INTRODUCTION

Water scarcity has emerged as one of the most pressing challenges of the 21st century. The rapid population growth, urbanization, industry, and climate change have intensified the demand for freshwater resources[13,24,44]. Over 66% of the global population encounters severe water scarcity for a minimum of one month annually, and projections indicate that numerous nations currently classified as water-secure may face water stress by 2030[33]. Consequently, preserving stable freshwater resources is both an environmental necessity and a significant socio-economic consideration. The necessity to tackle water constraint has accelerated the development of alternative, energy-efficient, and decentralized water harvesting systems. Traditional large-scale desalination facilities, encompassing thermal desalination and reverse osmosis technologies, have been widely deployed in dry coastal areas. Nonetheless, their substantial energy consumption, operational expenses, and environmental issues including brine disposal and greenhouse gas emissions continue to pose significant hurdles[4]. Consequently, atmospheric water collection methods have garnered heightened interest[5]. Solar-powered condensation systems employing double-slope surfaces have proven effective in extracting water from humid air under regulated conditions[43], while portable and foldable solar-driven devices have broadened the possibilities for dispersed applications[17]. Hygroscopic absorption-desorption systems utilizing materials like calcium chloride ($CaCl_2$) in conjunction with solar regeneration have been examined for the capture of ambient moisture[43,17]. Hybrid desalination approaches that combine fogging mechanisms with solar desalination systems have also been investigated to improve droplet production and distillate transport efficiency[2,3]. Emerging methodologies encompass membrane distillation, capacitive deionization, and nature-inspired water harvesting[49]. These techniques utilize engineered surface wettability and microstructural design to enhance efficient droplet nucleation and transport[22]. These solutions collectively demonstrate the extensive technological endeavor aimed at sustainable freshwater production within energy and environmental limitations.

While these methods are novel and often attractive in terms of collection efficiency, a major limitation arises from their reliance on energy supply and enhanced requirements on maintenance or infrastructure, rendering them unfeasible for deployment in remote places and rural or developing areas. Fog harvesting has recently been recognized as a passive and energy-efficient technique that operates with zero or minimum external power requirements. Fog collectors typically consist of mesh materials that capture airborne microdroplets carried by the wind, enabling the intercepted water to flow via gravity and collected in designated reservoirs[16]. The efficacy of fog harvesting has been demonstrated in multiple global locations, with effectiveness significantly influenced by climatic factors such as wind speed and liquid water content[1,31]. Natural fog consists of suspended water droplets, with the distribution of droplet sizes being a crucial element in aerodynamic interception. Field experiments indicate that fog droplet diameters vary from approximately 2 μm to 40 μm, typically exhibiting unimodal or bimodal distributions with peak values between 11 μm and 25 μm[32,15]. Due to the fact that droplet inertia is proportional to the cube of the diameter, droplets within this size range exhibit distinct behaviors in airflow across mesh materials, rendering them directly pertinent to fog harvesting efficiencies. Therefore, comprehending the aerodynamic characteristics of fog droplets within this realistic size spectrum is essential for improving fog collection. The efficiency of fog water collection is governed by multiple coupled physical processes, commonly interpreted through aerodynamic, deposition, and drainage mechanisms[14]. While most experimental studies report only the overall collection efficiency, such aggregate measures do not reveal the relative contributions of individual processes. In particular, capture efficiency defined as the combined effect of aerodynamic and deposition mechanisms provides a more direct measure of droplet–flow–mesh interaction prior to drainage[18,30]. This distinction is especially important when assessing the role of mesh geometry, since drainage behavior is additionally influenced by surface wettability and gravitational transport. Despite extensive research on fog harvesting materials and surface alterations, a comprehensive understanding of the impact of mesh shape on capture efficiency remains limited. Diverse mesh geometries, including rectangular, Raschel, and geometrically altered designs, have been evaluated in experimental configurations, nevertheless, the majority of these studies have just documented overall efficiency without investigating the aerodynamic characteristics of the process. The local flow acceleration, recirculation zones, and wake patterns generated by different mesh shapes can significantly influence droplet trajectories throughout the 2–40 μm diameter range. The effects of geometry-induced flow alteration can either facilitate or obstruct droplet impaction, depending on droplet inertia and streamline curvature. Notably, the coupled effects of mesh geometry, local flow structure, and droplet dynamics remain inadequately characterized.

Localized shear layers, wake recirculation, and velocity fluctuations around porous or bluff obstacles have been widely reported, where geometry-induced separation generates coherent unsteady flow structures and wake evolution strongly dependent on permeability and shape[46,50,11]. Recent studies further confirm that permeable obstacles can sustain jet-like flow structures, Kelvin–Helmholtz instabilities, and wake modulation depending on geometry and permeability[8]. Such hydrodynamic features strongly influence particle transport by modifying streamline curvature, residence time, and relative droplet–flow velocity. Their effect is particularly significant for droplets in the 2–40 µm range, where inertia-driven departure from streamlines governs interception efficiency. Consequently, geometrically different mesh configurations subjected to the same mean upstream flow can exhibit distinctly different capture behaviour.

The present work addresses this gap by introducing a spectral-dynamic framework that links mesh geometry, flow unsteadiness, and droplet capture efficiency. Unlike conventional approaches that focus primarily on mean-flow quantities or fluctuation intensity, the analysis is performed in the frequency domain ($f_d$ is the characteristic frequency) to identify characteristic timescales ($\tau_f$) associated with geometry-induced flow structures. By comparing these timescales with the droplet response time ($\tau_p$), the parameter, $\Pi = \tau_p/\tau_f = \tau_p f_d$, is defined. This parameter can be interpreted as a frequency-based analogue of the Stokes number: the flow timescale is obtained from the dominant spectral content rather than from a prescribed mean velocity and geometric length scale. Because $f_d$ is extracted from the simulated flow field, Π is data-informed rather than fully a priori predictive; predictive use would require independent estimation of $f_d$, for example from Strouhal-number correlations for a given geometry. The central hypothesis is that capture efficiency is enhanced when droplet response and flow unsteadiness are dynamically matched, rather than when fluctuations are simply strongest. This perspective provides a consistent physical basis for interpreting why geometries with moderate, spatially distributed fluctuations outperform those with weak or highly localized fluctuations.

The results show that mesh geometry controls the redistribution of spectral energy across pore and obstruction regions, thereby defining the effective flow timescale experienced by droplets. A physics-inspired correlation for capture efficiency is developed from the compatible droplet–flow response to provide a unified interpretation of the geometry-dependent trends. The framework is best viewed as a mechanistic design guide for fog-harvesting meshes, with broader optimization requiring additional geometries, velocities, and experimental validation.

# 2. Methodology

## 2.1 Eulerian framework

Fog mesh wire width was taken as the characteristic length scale ($l_f$) for our study, since the governing flow interactions occur locally around individual strands and pore openings. As the fog collector behaves as a porous array of repeated obstacles, the wire-scale dimension provides an appropriate basis for Reynolds number estimation and flow regime assessment[14]. Accordingly, the carrier phase is governed by the laminar, incompressible Navier–Stokes equations:

$$\nabla \cdot \mathrm{u} = 0 \qquad (1)$$

$$\frac{\partial \boldsymbol{u}}{\partial t} + \nabla(\boldsymbol{u}\,\boldsymbol{u}) = -\frac{1}{\rho}\nabla p + \nu\nabla^2 u + \boldsymbol{S}_{pf} \quad (2)$$

where $u$ denotes the fluid velocity, $p$ the pressure, $\rho$ the fluid density, ν the kinematic viscosity, and $S_{pf}$ represents the momentum source linked to interactions between dispersed Lagrangian droplets and the surrounding fluid (Appendix A1), primarily accounting for drag and other forces exerted by the droplets on the continuous phase.

## 2.2 Lagrangian framework

In the Lagrangian Particle Tracking (LPT) paradigm, liquid droplets are modeled as point particles that possess mass and momentum without occupying a finite physical volume. To save computational expenses, droplets with analogous characteristics—such as diameter, velocity, temperature, and thermophysical properties—are consolidated into computational entities known as parcels. Each parcel signifies a cluster of droplets with uniform characteristics and is monitored across the flow field during the simulation. At each time-step, the particle motion is revised by solving the Basset–Boussinesq–Oseen (BBO) equation[38], which dictates the translational dynamics of scattered particles:

$$\frac{dx_p}{dt} = \mathbf{u}_p \quad (3)$$

$$m_p \frac{d\boldsymbol{u}_p}{dt} = \boldsymbol{F}_{drag} + \boldsymbol{F}_G \quad (4)$$

where $\boldsymbol{x}_p$, $m_p$and $\boldsymbol{u}_p$ corresponds the position, mass, and velocity of each particle, respectively. $\boldsymbol{F}_D$ and $\boldsymbol{F}_G$ denotes the drag and gravitational (body) forces acting on the particles, and are computed as follows:

$$\boldsymbol{F}_D = C_D \frac{\pi D_p^2}{8}\, \rho g(\boldsymbol{u} - \boldsymbol{u}_p)\left|\boldsymbol{u} - \boldsymbol{u}_p\right| \quad (5)$$

$$\boldsymbol{F}_G = m_p g \quad (6)$$

where $D_p$ represents the particle diameter, and $C_D$ is the drag coefficient of a particle, calculated using the Schiller-Naumann equation[41]:

$$C_D = \begin{cases} \frac{24\left(1+0.15(Re_p)^{0.687}\right)}{Re_p} & Re_p \leq 1000 \\ 0.44 & Re_p > 1000 \end{cases} \quad (7)$$

The particle Reynold's number ($Re_p$) is defined as

$$Re_p = \rho D_p \frac{|\boldsymbol{u}-\boldsymbol{u}_p|}{\mu}\ , \quad (8)$$

where μ represents the dynamic viscosity of the continuum phase.

Carrier-phase fields ($u$, $\rho$, $\mu$) are interpolated to each particle location using trilinear interpolation. Particle trajectories are advanced with parcel sub-stepping so that the particle Courant number remains below the prescribed limit. Reaction forces from all parcels are accumulated and mapped to the corresponding host control volumes, enabling two-way momentum coupling between the dispersed droplets and the carrier gas. Upon collision with a solid boundary (i.e., a mesh fiber), a droplet is assumed to adhere to the surface, considering the hydrophilic nature of the metal mesh[19]. Literature suggests that, for hydrophilic surfaces, droplet bounce-off is not commonly observed at low impact Weber numbers[48]. Concurrently, a remarkably strong correlation was reported by Mundo et al (1995)[35] which showed that post-impact splashing is suppressed unless the impact parameter $K = Oh \times Re_p^{1.25}$ exceeds a threshold of 57.7 (the droplet Ohnesorge number, defined as $Oh = \mu/\sqrt{\rho\gamma D_d} = \sqrt{We}/Re_p$, relates the viscous force to inertial and surface tension forces). For the current range of operational conditions for fog droplets of $D_d \in$ (2–40 μm), advected in a stream velocity of $U$ =5 m/s the pertinent non-dimensional parameters $Re_p \in [10, 200]$, $We \in [0.715, 14.3]$, and an $Oh \in [0.019, 0.085]$, which clearly remains in the non-splashing and non-bouncing regime. This supports the simulation assumption that the fog droplets stick to the metal mesh fibers without bouncing or splashing. Accordingly, the Johnson-Kendall-Roberts (JKR) framework is used here as the physical motivation for this low-impact-energy adhesion limit; the simulations do not solve the full JKR contact-mechanics problem. Instead, they implement deterministic adhesion, i.e., no rebound or splashing after impact. The validity of this assumption is restricted to dilute-fog conditions involving micron-scale droplets and low impact Weber numbers. Under conditions involving larger droplets, higher impact velocities, substantial liquid-film accumulation, or strong aerodynamic loading, additional mechanisms such as rebound, splashing, film drainage, and secondary entrainment may become important. These processes can reduce the net water collection efficiency by returning previously captured liquid to the airflow or by altering the effective collector geometry. Such effects are not considered in the present study. Consequently, the reported efficiencies should be interpreted as capture efficiencies under low-Weber-number fog conditions. Extension of the framework to include probabilistic rebound, film evolution, drainage, and re-entrainment models represents an important direction for future work.

Reaction forces from all parcels are accumulated and mapped to the corresponding host control volumes, enabling two-way momentum coupling between the dispersed droplets and the carrier gas. The droplet loading corresponds to a representative fog liquid water content of 0.5 g/m$^3$, which yields a dispersed-phase volume fraction of approximately in order of $10^{-6}$ – $10^{-7}$, indicating a highly dilute flow regime. Consequently, momentum feedback from the dispersed phase to the carrier phase is expected to be weak; nevertheless, two-way coupling is retained through the Eulerian–Lagrangian source terms to ensure physical consistency.

### 2.3 Domain and boundary conditions

Figures 1 and 2 illustrate the computational domain used in this investigation. The domain features a prism of square cross-section measuring 180 $l_f$ × 180 $l_f$ in the x–z plane, and 300 $l_f$ in the streamwise direction (y-direction), with $l_f$ being the characteristic length of the fog mesh (defined as the representative geometric scale governing flow-mesh interaction, taken as the fiber diameter for cylindrical elements and an equivalent linear dimension for non-cylindrical mesh geometries). The characteristic length ($l_f$) is defined as the strand width normal to the incoming flow and is fixed at 0.2 mm for all geometries. This choice provides a consistent frontal obstruction scale for flow–mesh interaction and enables the influence of mesh topology to be isolated from variations in strand size. Furthermore, all geometries were constructed with the same shade coefficient (solid fraction) of 0.6 to ensure that differences in flow behavior and capture efficiency arise primarily from geometric arrangement rather than variations in blockage ratio. The fog mesh is located 75 $l_f$ downstream from the inlet boundary to provide adequate development of the entering droplet-laden flow before it interacts with the mesh. A downstream clearance of 225 $l_f$ separates the mesh from the outflow boundary, reducing potential boundary effects on droplet collision and wake formation around the fog mesh. A minimum lateral gap of 75 $l_f$ is preserved between the mesh and the domain side boundaries to prevent undesirable confinement effects. To facilitate equitable comparison among various geometrical configurations, the dimensions of the computational domain and the total fog mesh size are maintained consistently across all mesh shapes examined in the study. A consistent free-stream velocity of 5 m/s is specified at the inlet boundary. The fog mesh fibers are regarded as no-slip boundaries for the carrier phase. The lateral boundaries are positioned adequately distant from the mesh and are regarded as far-field boundaries to prevent the creation of false boundary layers along the domain edges. A constant pressure outlet condition is implemented at the outlet boundary.

A locally revised computational grid is implemented around the fog mesh to enhance the spatial resolution of velocity gradients and droplet-flow interactions, as illustrated in Fig. 2. In all configurations, the foundational computational grid comprises around 2.5 million cells, with fluctuations of about ±10% contingent upon the precise geometry of the fog mesh. An adaptive mesh refinement (AMR) technique is used to the base grid utilizing three refinement levels to identify areas with potentially significant velocity gradients. The refinement process is initiated when the velocity gradient magnitude is between 0.01 and 215, enabling the subdivision of cells along the interface until the maximum refinement level is attained or the velocity gradient is adequately resolved. The lower barrier of 0.01 is established after preliminary studies, which indicated that values below 0.01 do not yield discernible improvements in results, while raising the upper limit beyond 215 similarly shown no substantial enhancement in solution correctness. This localized refinement method allows for the use of a moderately sized base mesh of around 2.5 million cells while attaining a local resolution of approximately 100 µm in the refined interface region. Dynamic cell refinement is utilized in the simulation to precisely capture localized fluctuations in velocity and pressure as Lagrangian droplets traverse the domain. The solution dynamically refines areas of interest according to changing flow topologies. Consistent with Capecelatro's suggestions[9], the computational cell size is preserved at a minimum of twice the diameter of the Lagrangian droplet, even post-refinement. It is crucial to maintain a scale separation between the carrier phase grid and the dispersed phase particles to ensure the integrity of the Eulerian–Lagrangian framework, as Capecelatro[9] demonstrated that the computational mesh must be significantly larger than the particle diameter to yield physically meaningful results. In areas where high spatial resolution is unnecessary, the grid automatically defaults to its parent mesh structure, enhancing computing efficiency. To prevent sudden transitions that may create artificial velocity or pressure gradients, four buffer layers are implemented to ensure a steady transition between varying refinement levels. The dynamic refinement process increases the overall cell count from an initial base mesh of around 2.5 million to between 4 and 4.5 million, contingent upon the amount of injected droplets and their interactions with the fog mesh during the simulation. A maximum Courant–Friedrichs–Lewy (CFL) number of 0.5 is imposed during the computation, with the time step automatically modified to uphold this limitation.

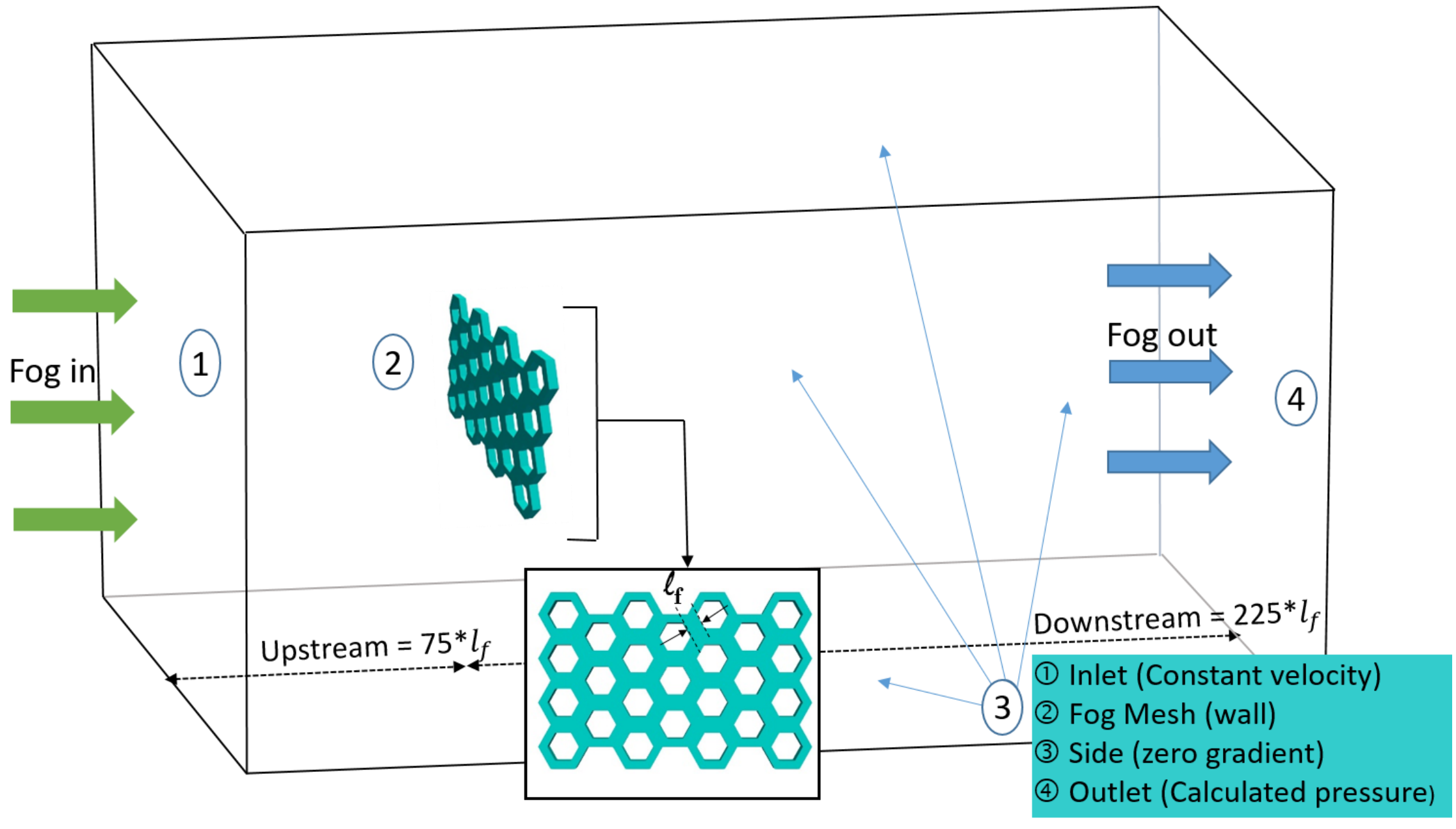

FIG 1. Schematic illustrating the computational domain(hexagon) of fog harvesting and the topology of fog mesh (the inset shows the mesh geometry). $l_f$ = Fog mesh characteristic length.

Time integration is executed with an implicit first-order backward Euler method with adaptive time stepping limited by CFL $\leq$ 0.5. Spatial discretization adheres to the conventional finite-volume approach utilized in OpenFOAM. The gradient and Laplacian terms are computed using the corrected Gauss-linear method, which corresponds to second-order central differencing for diffusion terms. Convective terms are discretized with a limited second-order vanLeer technique to ensure precision and mitigate spurious oscillations. The PIMPLE algorithm manages pressure–velocity coupling by employing multiple outer corrector iterations per time step (for instance, two outer loops with three pressure corrections and one non-orthogonal correction), thereby ensuring stability in the two-way momentum coupling between droplets and the carrier phase. Linear systems are resolved utilizing the GAMG solver for pressure and PBiCGStab for velocity. Absolute residual tolerances are established at $10^{-6}$, and are tightened to $10^{-8}$ in the final iteration. Convergence is deemed accomplished when all residuals meet these thresholds and the instantaneous continuity error is less than $10^{-8}$.

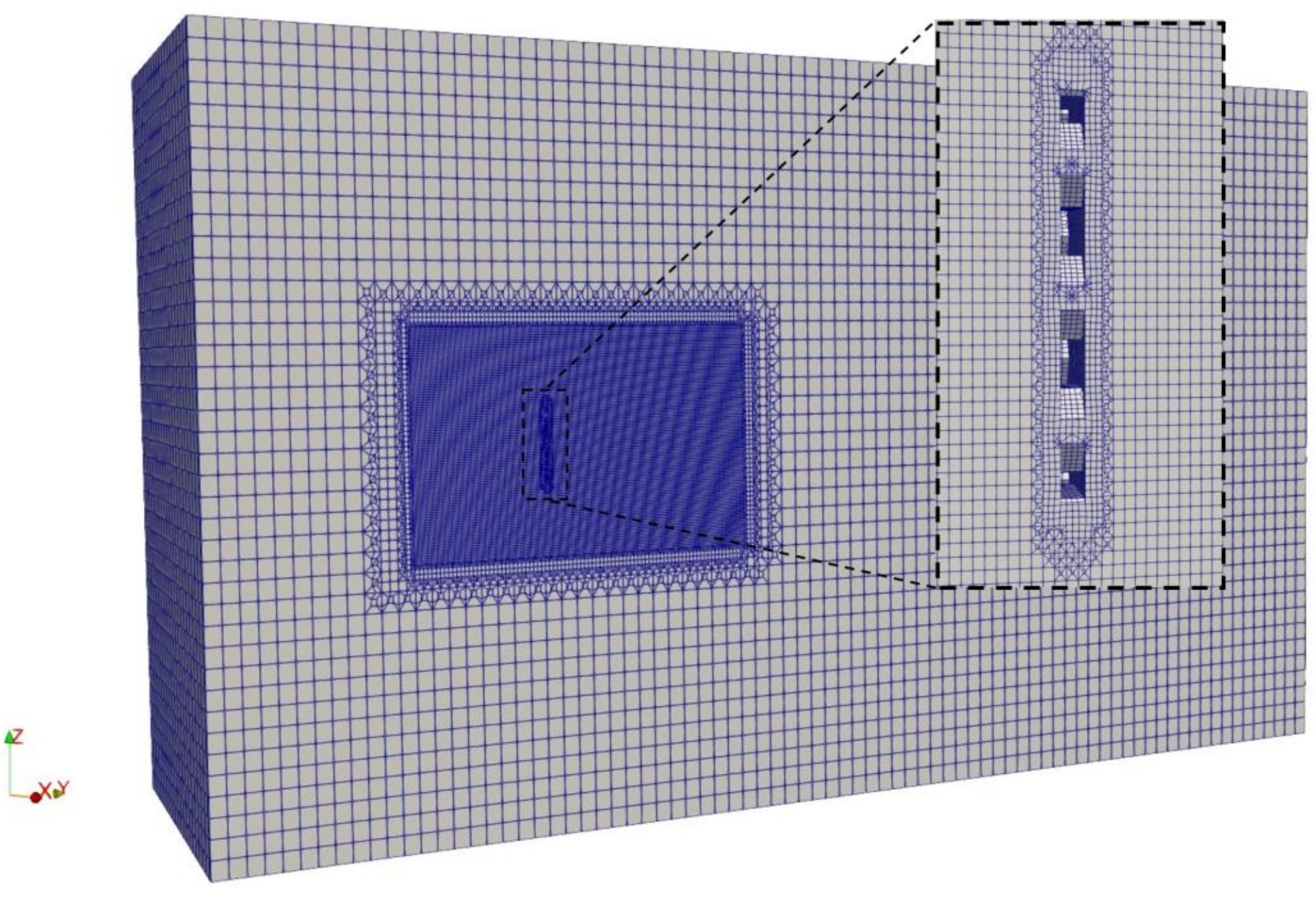

(a)

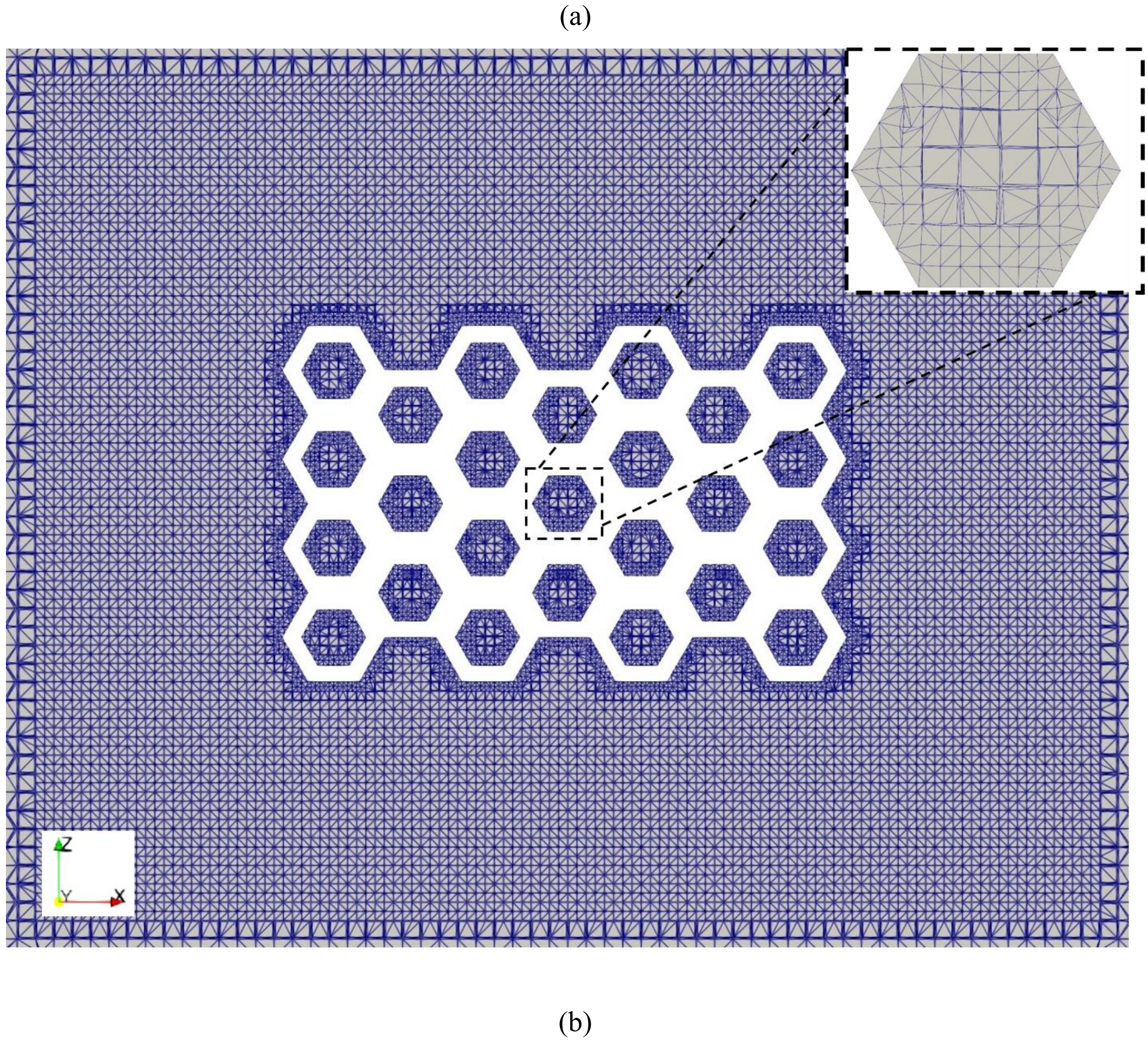

(b)

FIG 2. Computational grid generated using blockMesh (OpenFOAM). (a) Overview of the whole domain (inset shows the grid refinement around the fog-harvesting mesh). (b) Frontal view of the fog mesh (fine-tuned using snappyHexMesh).

## 2.4 Framework for droplet-flow interaction and spectral analysis

This section outlines the integrated framework used to model droplet-flow interaction and quantify capture efficiency in the presence of geometry-induced unsteadiness. The approach consists of three key components: (i) representation of the inlet fog droplet population using a physically consistent size distribution, (ii) Lagrangian modeling of droplet transport and capture within the flow field, and (iii) frequency-domain analysis of velocity fluctuations to characterize the unsteady flow structures governing droplet dynamics. Detailed formulations, numerical procedures, and parameter selections associated with these components are provided in the Appendix A1-A3.

Fog droplets in natural atmospheric conditions typically span diameters from approximately 2 μm to 40 μm, as reported in field observations[22,37,40]. To represent this polydispersity, a two-parameter Rosin-Rammler distribution is employed to define the inlet droplet size distribution. This formulation enables consistent representation of macroscopic properties such as liquid water content and effective droplet size, while allowing flexible parameter calibration[6,36]. The selected distribution parameters are guided by in-situ measurements from the SIRTA observatory, which indicate unimodal droplet spectra with peak concentrations in the range of 11–25 μm[32]. This approach ensures that the injected droplet population remains both physically representative and compatible with the Eulerian-Lagrangian framework implemented in OpenFOAM. A detailed description of the distribution formulation and parameter selection is provided in Appendix A2.

The dispersed phase transport is modeled using Lagrangian particle tracking, with droplets injected at the inlet and their trajectories resolved under the influence of the carrier-phase flow. The standard fog-harvesting collection process is commonly decomposed into aerodynamic, deposition, and drainage mechanisms[14]. In the present study, the emphasis is on capture efficiency, which combines the aerodynamic interception and deposition components. Droplets that contact mesh fibers are assumed to adhere deterministically, so rebound and secondary breakup are neglected. Under this assumption, capture efficiency is evaluated as the ratio of droplets deposited on the mesh to droplets entering the corresponding projected area at the inlet plane① of the computational domain Fig. 2(a). Drainage of deposited liquid over the mesh-fiber surface is not modeled because it involves additional multiscale surface and film dynamics beyond the scope of the present aerodynamic analysis. Further details of the numerical implementation and efficiency formulation are provided in Appendices A1 and A3.

To characterize the unsteady flow structures governing droplet transport, frequency-domain analysis is performed on the carrier-phase velocity field using Fast Fourier Transform (FFT) and Power Spectral Density (PSD) techniques. Spectral methods are widely employed in fluid dynamics to identify dominant flow structures and quantify energy distribution across frequency scales[39,45]. Velocity signals are sampled at multiple probe locations distributed across key flow regions, including upstream, near-mesh, and wake zones, enabling spatial characterization of flow unsteadiness. The analysis is conducted on the fluctuating component of the streamwise velocity, ensuring that the extracted spectra reflect only dynamic variations in the flow. The interaction between droplets and the carrier flow is quantified by comparing the spectral energy of droplet-laden and single-phase flows. The spectral gain provides a measure of droplet-induced amplification of velocity fluctuations, while the spectral difference quantifies the redistribution of energy across frequencies. Together, these metrics enable identification of frequency bands where droplet-flow coupling is most significant. Complete details of the signal processing methodology and spectral formulations are provided in Appendix A4.

## 2.5 Solver validation

Accuracy of the solver has been validated against numerical predictions of Chen[10], whose trajectory-based analysis demonstrated enhanced concordance with Langmuir's experimental data[28]. The validation study involves simulating droplet motion around a cylindrical obstruction within a three-dimensional computational domain of dimensions $30D \times 20D \times 20D$ in the $x$, $y$, and $z$ axes, where $D$ represents the cylinder diameter. The computational mesh is significantly improved near the cylinder surface to precisely capture the flow field and droplet trajectories, resulting in about 1.212 million cells within the domain. The primary simulation parameters and fluid characteristics utilized in the validation scenarios are described in Tables 1 and 2, respectively. In the initial validation setup (Case 1), the cylinder diameter is set at 5 mm. A second arrangement featuring a decreased cylinder diameter of 0.3 mm is also evaluated. In this smaller cylinder, the input velocity and droplet size are modified to ensure that the resulting Stokes number, defined as

$$St = \frac{\rho_d D_d^2 U}{18\mu_a l_f} \tag{9}$$

(where $U$ denotes the upstream flow velocity, $\mu_a$ the air viscosity, $\rho_d$ represent the droplet density, $l_f$ represents the characteristic length which in this case is the cylinder diameter ($D$), and $D_d$ denotes the droplet diameter) used in the simulations corresponds with the reference case documented by Chen[10]. Figure 3(a) illustrates a cross-sectional representation of the computational domain employed for code-validation.

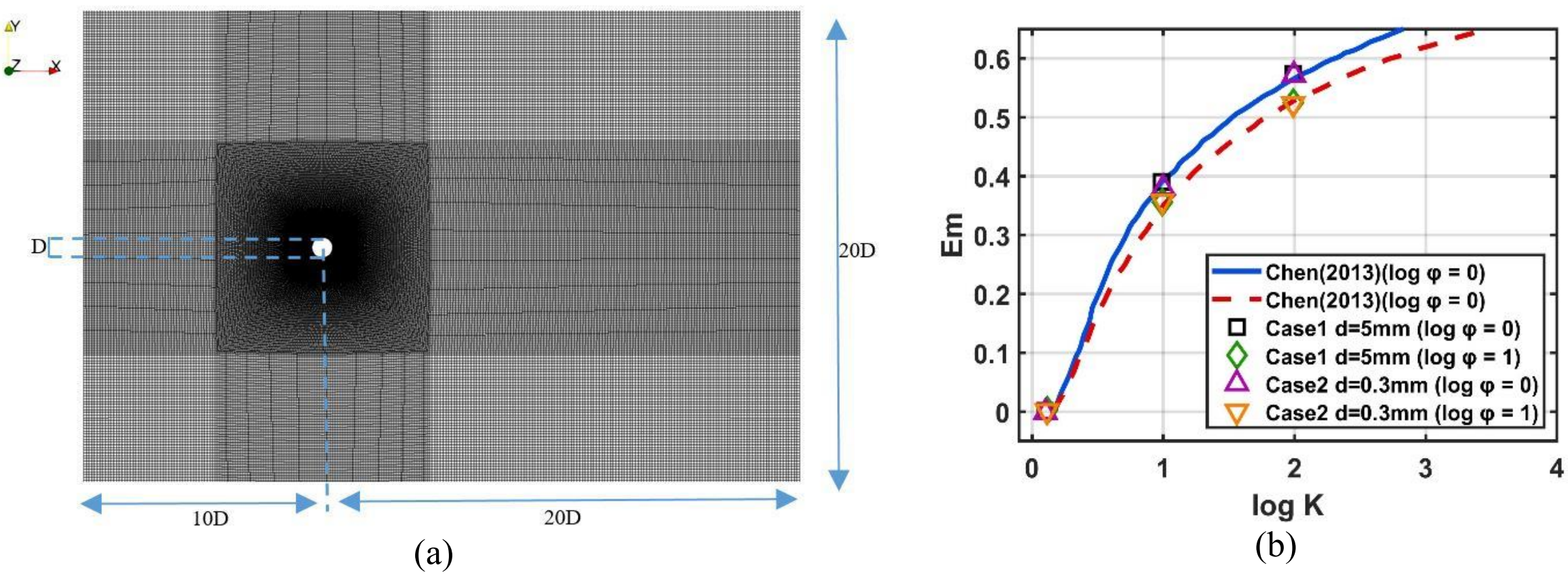

FIG 3. (a) Computational domain and grid distribution adopted for case-1 (D=5mm) and case-2 (D=0.3mm) (b) Comparison of collision efficiency of both cases and numerical report Chen et al[10]

The anticipated capture efficiencies derived from the current simulations are juxtaposed with the overall collision efficiency (*Em*) of a single cylinder as documented by Chen[10]. In Chen's study, *Em* denotes the proportion of droplets that impact the cylindrical collector, and the trajectory-based findings demonstrated a superior alignment with Langmuir's experimental data compared to previous analytical models. Within Langmuir's paradigm[26], the parameter $K$ denotes an inertial-impaction parameter that integrates droplet inertia with distinctive flow scales, while φ functions as an auxiliary parameter utilized alongside $K$ to correlate the collision efficiency. These parameters are delineated as

$$K = \frac{D_d^2 \rho_d U}{9 \mu_a D} \tag{10}$$

$$\varphi = \frac{\mathrm{Re}^2}{K} = \frac{9 \rho_a^2 U D}{\mu_a \rho_d} \tag{11}$$

while $\rho_a$ represent the density of air.

*Em* measures the ratio of droplets that collide with the cylindrical element to the total number of droplets starting on trajectories that could potentially contact the cylinder. It quantifies the proportion of entering droplets whose initial locations and velocities permit possible collisions with the cylinder and that finally make contact with the surface. The current wall-interaction model mandates deterministic adhesion upon impact (i.e., droplets stick to the surface without rebounding), so the collision efficiency *Em* reported by Chen[10] directly correlates with the capture efficiency anticipated in the present simulations. From a physical standpoint, droplets first adhere to the surrounding streamlines as the airflow accelerates and curves around the cylinder. Owing to their limited inertia, certain droplets diverge from the curved streamlines and impact the surface, but others are transported past the obstruction by the flow. Figure 3(b) juxtaposes the anticipated capture efficiency ($\eta_{cap}$) with Chen's *Em* for both cylinder diameters examined in this validation exercise.

**TABLE 1. Test conditions used in validation at reference atmospheric condition (28 $^oC$ and 1 bar).**

| Variables | Case 1(5 mm) | | | Case 2 (0.3mm) | | |
|---|---|---|---|---|---|---|
| | Case 1a | Case 1b | Case 1c | Case 2a | Case 2b | Case 2c |
| *logK* | 0.1 | 1 | 2 | 0.1 | 1 | 2 |
| *log*φ | 0 | 0 | 0 | 1 | 1 | 1 |

**TABLE 2. Summary of fluid properties taken at atmospheric condition (28 $^oC$ and 1 bar)**

| *Fluid properties* | *Values taken during simulation* |
|---|---|
| $\rho_d$ (droplet density, kg/m$^3$) | 996.1 |
| $\rho_a$ (air density, kg/m$^3$) | 1.172 |
| $\mu_d$ (droplet dynamic viscosity, N.s/m$^2$) | $8.318\times10^{-4}$ |
| $\mu_a$ (air dynamic viscosity, N.s/m$^2$) | $1.851\times10^{-5}$ |

Among the simulated cases, Cases 1c and 2b (see Table 1) exhibit the largest deviations from the reference values, with differences of 1.4% and 3.2%, respectively. The results for the 5 mm cylinder (Case 1) show very close agreement with the reference data, while the 0.3 mm cylinder (Case 2) demonstrates satisfactory correspondence across the examined range of *St*. The use of a single-cylinder geometry for validation is standard practice in the fog-harvesting literature[14,30] as it provides a well-characterized benchmark with analytical solutions for inertial impaction efficiency. While mesh arrays introduce wake interactions that are absent from the single-cylinder case, the fundamental droplet capture physics validated here – inertial impaction, drag, and wall adhesion are geometry-independent and remain relevant even in the mesh-array simulations. Wake interactions between adjacent fibers in the full mesh are explicitly resolved by the flow solver (not modeled analytically), and their spectral signatures are captured by the frequency-domain analysis in Section 3.3. The validated solver is therefore considered appropriate for the mesh-array simulations presented in this study.

## 2.6 Grid-Independency test

A grid-independence analysis has been conducted to guarantee that the numerical predictions are unaffected by the computational mesh resolution. Given that all fog mesh configurations examined in this study possess similar computational domain dimensions and vary solely in the shape of the centrally positioned fog mesh, the grid-independence analysis is conducted using the trapezoidal mesh configuration as a reference example. This method guarantees that the chosen mesh resolution is optimal. Three distinct base mesh resolutions are analyzed: coarse, medium, and fine meshes. The coarse mesh comprised roughly 1.6 million cells, which escalated to approximately 3.0–3.3 million cells following the implementation of adaptive mesh refinement (AMR). The medium mesh comprised roughly 2.5 million base cells, with Adaptive Mesh Refinement augmenting the overall cell count to approximately 4.0–4.5 million. The optimal mesh comprises roughly 3.2 million base cells, which expands to approximately 5.2–5.6 million cells upon refining.

The AMR technique is uniformly implemented across all computational -mesh layers utilizing consistent refining criteria derived from local velocity gradients. This method adaptively enhances the mesh resolution in areas with significant velocity gradients, especially adjacent to the fog mesh fibers where droplet-flow interaction takes place, while preserving a coarser mesh in parts with relatively uniform flow. Thus, the refinement procedure guarantees precise delineation of the flow features that influence droplet trajectories and capture behavior.

The capture efficiency ($\eta_{cap}$) serves as the principal metric for assessing the grid convergence. Table 3 delineates the mesh resolutions alongside the associated capture efficiency values for the three mesh levels.

**TABLE 3. Grid Independency test**

| *Grid Level* | *Base cells (Million)* | *Final Cells after AMR (Million)* | *Capture Efficiency* |
|---|---|---|---|
| *Coarse* | 1.6 | 3.0-3.3 | 0.5488 |
| ***Medium*** | **2.5** | **4.0-4.5** | **0.5562** |
| *Fine* | 3.2 | 5.2-5.6 | 0.5596 |

Table 3 suggests that the anticipated capture efficiency marginally improves with grid refinement. The percentage difference between the medium and fine grid findings is roughly 0.61%, suggesting that more refinement yields relatively minimal alterations in the expected capture efficiency. This trend indicates that the numerical solution has successfully attained grid-independent conditions. The medium grid resolution was used for all simulations in this study based on the analysis conducted offering an optimal compromise between computing precision and expense, while guaranteeing that the anticipated capture efficiencies are unaffected by grid resolution.

# 3. Results and Discussion

The present analysis investigates droplet–flow interaction across five representative mesh geometries—triangular, square, trapezoidal, hexagonal, and vertical (harp-like) selected to systematically vary pore topology, streamline curvature, and blockage characteristics. These geometries span configurations commonly used in fog-harvesting systems as well as idealized shapes that isolate geometric effects on flow restructuring. All simulations are conducted at a fixed inlet velocity of 5 m/s, representative of moderate wind conditions typical of fog-harvesting environments (note that the single-velocity restriction limits direct verification of the framework's universality; multi-velocity simulations are recommended as essential future work), with the droplet size distribution ranging from 2 to 40 μm based on field observations. The inlet fog loading is prescribed through a Rosin–Rammler distribution (Appendix A3), ensuring realistic representation of both number density and mass contribution of droplets. The results are presented to first establish the geometry-induced restructuring of the flow field, followed by its influence on capture efficiency, and subsequently interpreted through spectral analysis and timescale compatibility between droplet response and flow timescales.

## 3.1 Geometry-induced restructuring of the near-mesh flow and probe placement

The interaction between the incoming flow and the mesh produces a strongly heterogeneous near-field characterized by local acceleration, deceleration, and shear-layer formation. Although the inlet flow is uniform, the presence of the mesh partitions the domain into two dynamically distinct regions: (i) an obstruction-dominated region adjacent to the solid elements, where flow decelerates and separates, and (ii) a pore-dominated region, where flow accelerates through the mesh openings. This decomposition forms the basis for understanding droplet behavior, as the extent of fog-droplet capture depends on how momentum and unsteadiness are redistributed between these regions.

Figure 4 illustrates the near-field flow structure for the triangular mesh, highlighting the spatial variation of velocity in planes intersecting the mesh. The flow exhibits pronounced streamline deflection upstream of the solid elements, accompanied by the formation of distributed shear layers that extend across the pore region. Unlike geometries that concentrate acceleration within narrow channels, the triangular configuration induces spatially distributed flow perturbations, indicating that the flow is modified over a broader region rather than through isolated high-gradient zones. This behavior is important because droplet capture is governed not only by local velocity magnitude but by the extent and persistence of flow restructuring. Distributed shear layers increase the probability of droplet deviation from streamlines while simultaneously prolonging residence time in the near-mesh region. In contrast, geometries that produce highly localized acceleration can reduce interaction time by rapidly advecting droplets through the pore.

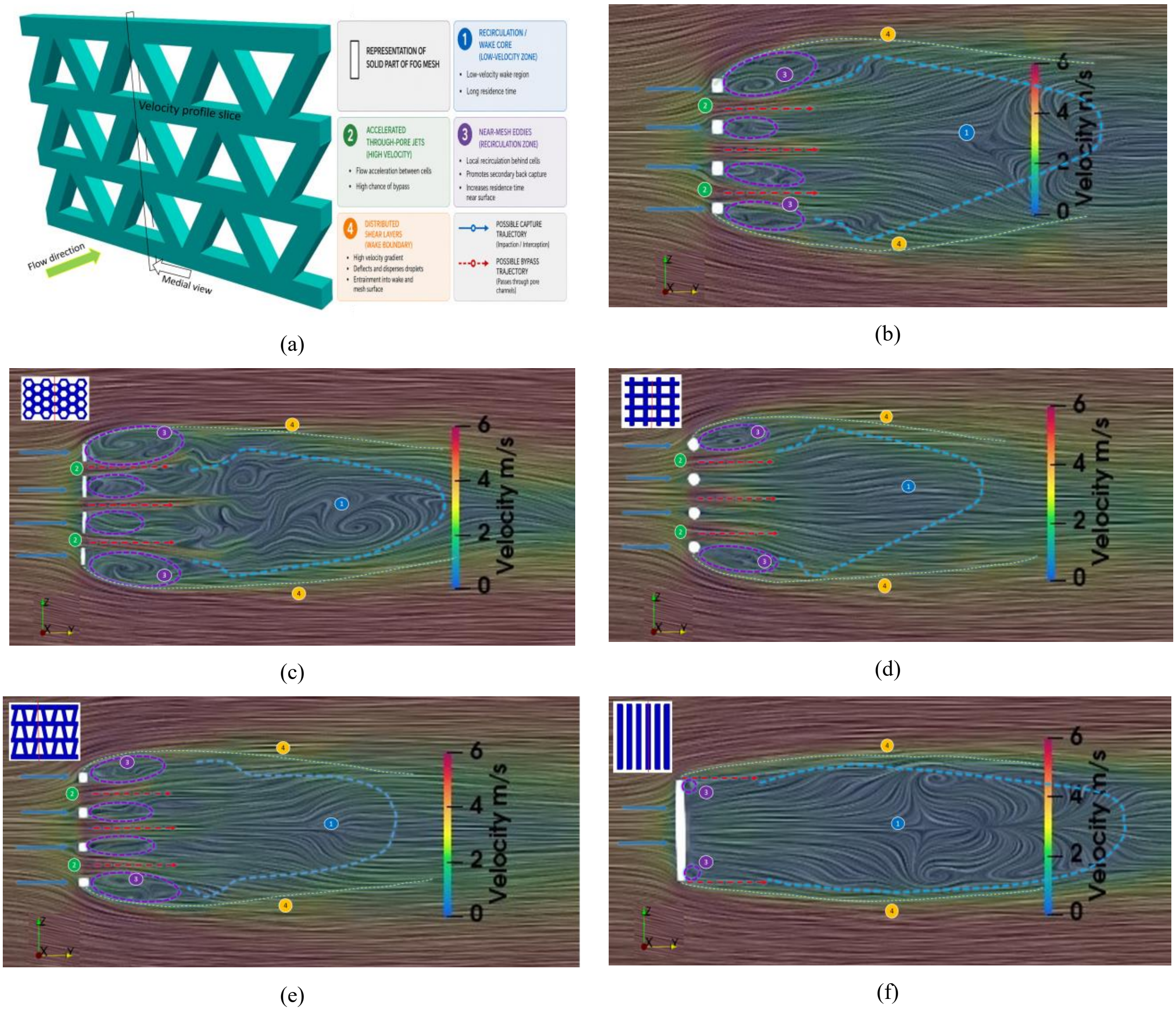

Fig 4. (a) Enlarged geometric representation and velocity profile of the triangular mesh, illustrating the flow-structure classification adopted in this study, including the wake core (Region 1), accelerated through-pore jets (Region 2), near-mesh eddies (Region 3), distributed shear layers (Region 4), and representative droplet capture and bypass trajectories. The notation, symbols, and region annotations introduced the right hand side panel are consistently employed throughout the remaining figures. (b) Velocity profile at the representative slice of the rectangular mesh. (c–f) Geometric configurations (the top-left inset cross-sectional views) and the corresponding velocity profiles of the (c) hexagonal, (d) square, (e) trapezoidal, and (f) vertical meshes, respectively.

The annotated flow sections further clarify the dominant transport mechanisms. Section 1 corresponds to the central low-velocity recirculation or wake core formed downstream of the mesh, where flow reversal and reduced momentum indicate an extended interaction region. Section 2 identifies the accelerated bypass streams passing through the open pore gaps, where droplets that remain strongly coupled to the carrier phase are more likely to traverse the mesh without interception. Section 3 represents localized near-mesh eddies generated immediately behind the solid strands, which can redirect nearby droplets toward adjacent collection surfaces and promote additional interception opportunities. Section 4 denotes the outer distributed shear layers separating the wake from the free stream, where sustained velocity gradients enhance lateral droplet migration and streamline deviation. Together, these regions demonstrate that capture efficiency depends on the combined influence of pore-scale acceleration, wake retention, and shear-induced trajectory modification rather than on a single local velocity feature.

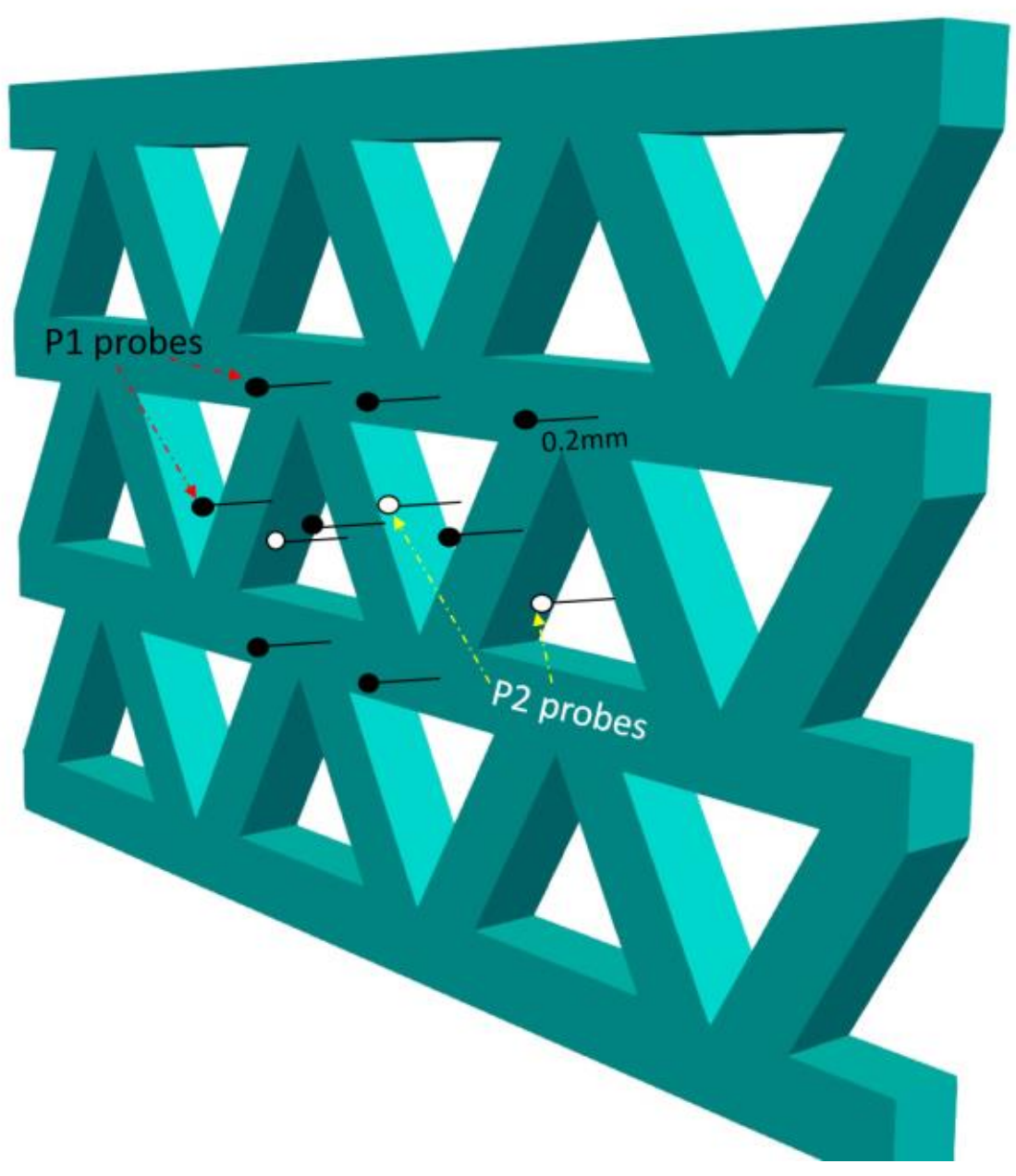

Fig 5. Probe location selected in order to address the underlying physics at local sections probes at the vicinity of fogmesh only with a distance of 0.02mm upstream under high pressure gradient zone. This is the zone where coupling of droplet and flow impacts the most at bypassing the fogmesh or adhering to it.

For a quantitative description of localized flow structures, multiple probes are placed at strategic near-mesh locations. Because the near-field contains both pore-accelerated flow and obstruction-driven deflection, probes are positioned to sample the open-pore region and the solid-strand region separately. For the triangular mesh shown in Fig. 5, P1 denotes the obstruction probes located near solid elements, whereas P2 denotes the pore probes located within the mesh openings. The P1 probes therefore quantify velocity gradients, separation, and deflection near fibers, while the P2 probes quantify accelerated bypass through pores. These complementary locations enable comparison of droplet-laden flow (DF) and single-phase flow-only cases (OF), so that P1-DF denotes the droplet-laden velocity signal at an obstruction probe.

## 3.2 Inertial transition and scaling of capture efficiency

Figure 6 presents the variation of capture efficiency with droplet diameter for all mesh geometries. It should be noted that the efficiencies reported here correspond to capture efficiency ($\eta_{cap}$), which quantifies the fraction of droplets intercepted and deposited on the mesh surface. In contrast, field studies commonly report overall collection efficiency ($\eta_{tot} = \eta_{cap} \times \eta_{dr}$), which additionally incorporates drainage and liquid-removal processes. While single-layer fog collectors often exhibit overall collection efficiencies in the range of 5–15%, optimized mesh designs[42] and industrial fog-harvesting environments[20] have reported efficiencies approaching 40%. Furthermore, drainage efficiency ($\eta_{dr}$) is strongly dependent on mesh geometry, surface wettability, and liquid loading, and has been reported to vary between approximately 45% and 95%[19]. Because drainage and liquid transport are not modeled in the present simulations, the results should be interpreted as idealized capture efficiencies rather than direct predictions of field-scale collection performance.

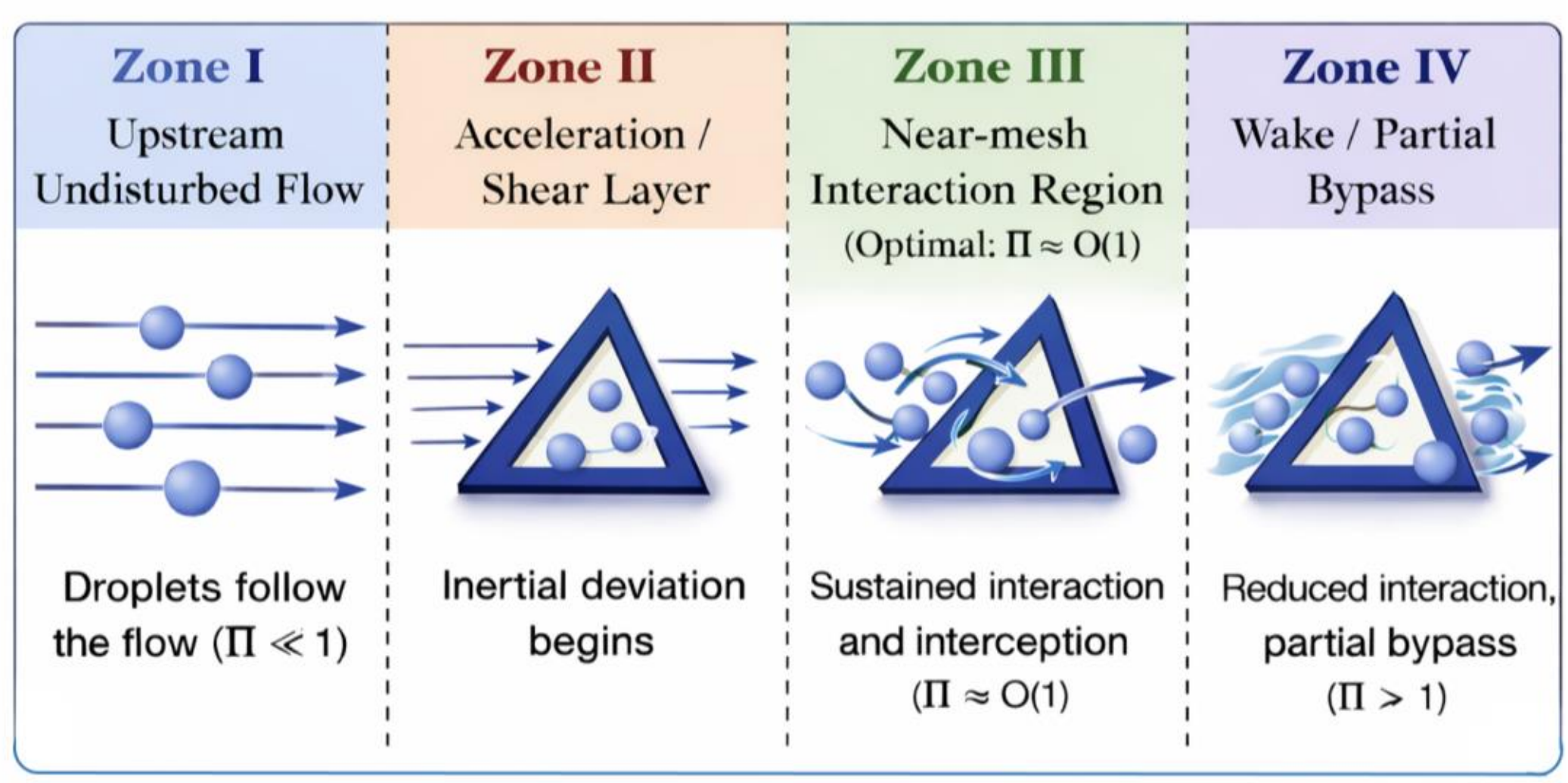

(a)

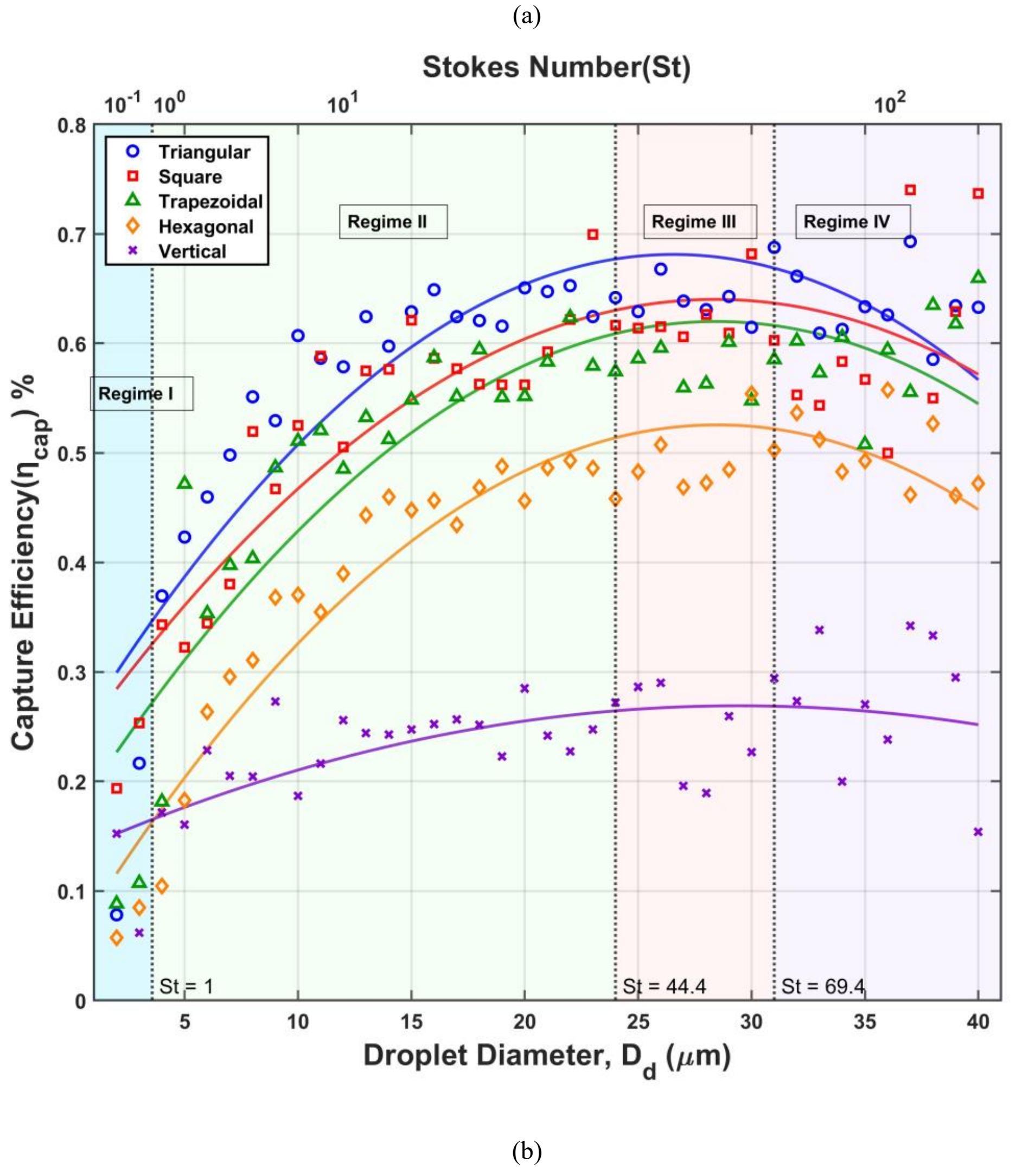

(b)

Fig 6. a) Schematic representation of droplet transport and interaction across different flow regions in a triangular fog mesh, illustrating upstream undisturbed flow (Zone I), onset of inertial deviation in the shear layer (Zone II), enhanced near-mesh interaction and interception (Zone III), and wake-induced partial bypass downstream of the mesh (Zone IV), b) Variation of droplet capture efficiency with droplet diameter for different mesh geometries at an inlet airflow velocity of 5 m/s.

Capture efficiency increases monotonically with droplet diameter over the range considered, reflecting the increasing importance of inertial effects on droplet interception. However, the increase in capture efficiency is not uniform across geometries, and the separation between the curves indicates that inertia alone does not determine capture performance. Instead, the efficiency reflects a combined effect of droplet inertia and geometry-induced flow restructuring. To facilitate interpretation of the results, the flow–droplet interaction is described using the schematic shown in Fig. 6(a), which is divided into four characteristic zones. In Zone I (upstream regime), droplets are strongly coupled to the flow and exhibit negligible interception. In Zone II (acceleration and shear-layer region), the onset of inertial deviation occurs as droplets begin to respond to geometry-induced flow gradients. The most significant interaction takes place in Zone III (near-mesh interaction region), where droplets experience sustained forcing due to local flow unsteadiness. In Zone IV (wake region), further increase in inertia yields diminishing returns, as droplets traverse the mesh with reduced interaction time, leading to partial saturation or bypass.

The mesh geometry-dependent ranking observed in Fig. 6(b) (triangular > square > trapezoidal > hexagonal > vertical) can be interpreted through the zonal framework introduced in Fig. 6(a), particularly in terms of how effectively each configuration sustains interaction within the Zone III. The triangular mesh exhibits the highest efficiency as it promotes extended droplet residence and sustained interaction in the near-mesh region, while avoiding excessive acceleration that would otherwise induce premature bypass. The square mesh – albeit generating strong local forcing – shows slightly reduced efficiency due to localized acceleration through the pores, which shortens interaction time. The trapezoidal configuration represents an intermediate case, whereas the hexagonal and vertical (harp) geometries perform poorly due to weaker or less coherent flow perturbations, limiting effective droplet-mesh interaction.

This behaviour is governed by the balance between droplet inertia and flow coupling, as reflected in the regime classification based on Stokes number. In Regime I ($St < 1$), droplets remain strongly coupled to the flow and follow streamlines, resulting in negligible interception. As inertia increases in Regime II, droplets begin to deviate from streamlines due to geometry-induced flow gradients, leading to a sharp increase in capture efficiency. Regime III represents the optimal scenario ($44.4 < St < 69.4$) where droplets possess sufficient inertia to interact effectively with the mesh, while still responding to local flow unsteadiness, resulting in maximum capture with a tendency toward saturation. In Regime IV ($St > 69.4$), further increases in inertia reduce droplet responsiveness to flow perturbations, leading to shortened interaction times, increased bypass, and a decline in capture efficiency.

Taken together, these results highlight that it is this balance between droplet inertia and flow coupling — rather than inertia in isolation — that governs capture performance, requiring droplets to experience sustained and dynamically compatible forcing within the near-mesh region. The observed trend of capture efficiency in Fig. 6(b) provides the basis for the spectral analysis presented in the subsequent section (Section 3.3), where the flow is examined in the frequency domain to identify the characteristic timescales governing droplet–flow interaction. The role of this dynamic compatibility is further formalized in Section 3.4 through the introduction of parameter $\Pi$, which offers a unified framework for interpreting the efficiency trends observed across different geometries and flow conditions.

### 3.3 Spectral Analysis of Droplet-Flow Interaction

To elucidate the mechanisms underlying the geometry-dependent trends in capture efficiency (Fig. 6), the droplet–flow interaction is examined in the frequency domain. While Section 3.2 established the macroscopic dependence of capture efficiency on droplet inertia, the present analysis focuses on how mesh geometry redistributes fluctuation energy across spatial regions and frequency scales, thereby modulating droplet trajectories and residence time. To avoid repetition, the detailed spectral characteristics are first discussed for the triangular mesh, which exhibits the highest capture efficiency. The behaviors for other geometries are subsequently interpreted in a comparative sense using summarized metrics (Table 4), with full spectra provided in Appendix A5-A8.

A. Spectral response for the triangular mesh

Figure 7 presents the frequency-domain diagnostics for the triangular mesh, including the Fourier spectrum, power spectral density (PSD), droplet-induced spectral variation (ΔPSD), and spectral gain ($G_1$). The analysis focuses on the transverse velocity component, which plays a dominant role in droplet deflection and interception. A defining feature of the triangular geometry is the broadband distribution of spectral energy, extending from low to intermediate frequencies without the emergence of isolated dominant peaks (Fig. 7(a,b)). Both probe locations, representing the obstruction (P1) and pore (P2) regions—exhibit comparable spectral magnitudes over a wide frequency range. This indicates that the geometry does not concentrate unsteadiness in a localized region but instead distributes it across the near-mesh flow. The PSD further confirms that droplet-laden flow enhances fluctuation energy primarily in the low-to-intermediate frequency band, corresponding to large- and intermediate-scale flow structures. Importantly, the energy levels at P1 and P2 remain of the same order, implying that neither the solid region nor the pore region dominates the dynamics. Instead, the flow exhibits spatially distributed unsteadiness, which is favorable for sustained droplet–flow interaction.

This interpretation is reinforced by the ΔPSD distribution (Fig. 7(c)), which shows a consistently positive and spatially uniform energy increment across frequencies. Unlike geometries that exhibit localized energy injection, the triangular mesh promotes a distributed enhancement of fluctuation energy, suggesting that droplets interact continuously with the flow rather than intermittently. It should be noted that at typical fog liquid water contents (0.5 $g/m^3$), the dispersed-phase volume fraction is of order $10^{-7}$–$10^{-6}$, so the two-way momentum coupling between the droplet and carrier phases is inherently weak[21]. Consistent with this weak coupling, the ΔPSD differences between DF and OF cases are therefore small in absolute magnitude, which should be interpreted as indicative trends rather than large-amplitude flow modifications. The primary value of the spectral framework lies in characterizing the geometry-driven flow timescales from the single-phase (OF) spectrum, against which the droplet response times are compared.

The spectral gain $G_1$ (Fig. 7(d)) remains moderate and broadband, with values close to unity over most of the spectrum and only mild amplification in specific bands. This behavior indicates that droplet-induced modulation of the flow is persistent but not excessive. In physical terms, the triangular geometry avoids both extremities: it neither under-excites the flow (leading to weak interaction) nor over-amplifies it (leading to rapid droplet bypass). Taken together, these observations indicate that the triangular mesh produces a balanced spectral environment, characterized by: (i) distributed energy across pore and obstruction regions, (ii) sustained low-to-intermediate frequency content, and (iii) moderate amplification without sharp localization. Such a configuration is conducive to prolonged droplet residence time and repeated interaction with the mesh, thereby enhancing capture efficiency.

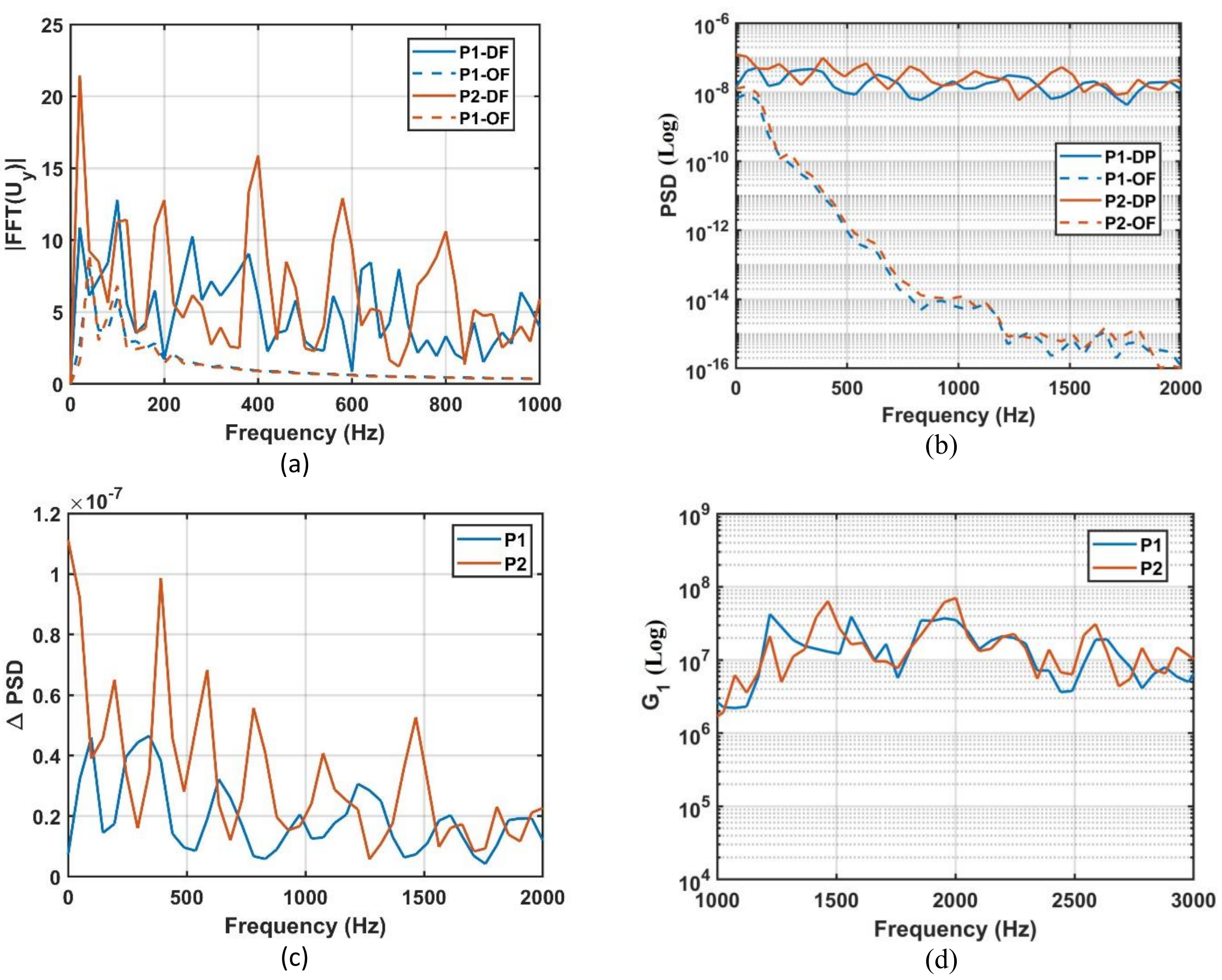

Fig. 7 Frequency-domain characterization of droplet–flow interaction for the triangular mesh: (a) Fourier spectrum of transverse velocity fluctuations, (b) power spectral density (PSD) comparing droplet-laden (DF) and droplet-free (OF) cases, (c) droplet-induced spectral variation ΔPSD, and (d) spectral gain $G_1 = S_1/S_2$. Results are shown for probes located in the obstruction (P1) and pore (P2) regions (see Fig 5). The triangular geometry exhibits broadband energy distribution and moderate amplification, indicative of sustained and spatially distributed droplet–flow coupling.

B. Comparative interpretation of spectral characteristics across different mesh geometries

The spectral characteristics of the remaining geometries are summarized in Table 4, with detailed plots provided in Fig. 4. The comparison reveals that the differences in capture efficiency (Fig. 6) can be directly interpreted in terms of how each geometry redistributes spectral energy. The square mesh, although highly efficient, exhibits strong and localized spectral amplification, particularly within the pore region. The presence of sharp peaks in both PSD and $G_1$ indicates that fluctuation energy is concentrated within narrow frequency bands. While such amplification enhances local droplet deflection, it also promotes rapid acceleration through the pores, thereby increasing the likelihood of bypass. This explains why the square geometry performs well but does not surpass the triangular configuration.

The trapezoidal mesh represents an intermediate case, characterized by weakly distributed low-frequency energy and near-unity spectral gain. The absence of strong amplification or clear spatial dominance suggests that droplet–flow interaction is present but not optimally sustained. As a result, the capture efficiency remains moderate. In contrast, the hexagonal mesh exhibits a spectrally smooth and low-energy response, with minimal droplet-induced enhancement. The lack of sufficient fluctuation energy implies that droplets remain largely entrained within the carrier flow, resulting in limited inertial deviation and reduced capture. The vertical (harp) configuration displays a markedly different behavior, characterized by intermittent and frequency-localized spectral peaks. Although high instantaneous amplification is observed at specific frequencies, the absence of sustained broadband energy leads to incoherent forcing. Such intermittency is unfavorable for droplet capture, which requires persistent interaction over timescales comparable to the droplet response time.

**TABLE 4.** Summary of spectral characteristics for different mesh geometries and their physical implications for droplet capture.

| ***Geometry*** | ***Dominant frequency range*** | ***Energy distribution (P1 vs P2)*** | ***ΔPSD behavior*** | ***G1 behavior*** | ***Physical interpretation*** |
|---|---|---|---|---|---|
| *Triangular (Fig 7)* | *Broadband (low–mid)* | *Balanced (P1 ≈ P2)* | *Distributed, positive* | *Moderate, broadband* | *Sustained interaction; optimal capture* |
| *Square (Fig–A8)* | *Narrow peaks (mid-frequency)* | *Pore-dominated* | *Localized high* | *Strong peaks* | *Over-amplification; partial bypass* |
| *Trapezoidal (Fig–A7)* | *Low-frequency dominant* | *Nearly balanced* | *Weak, low-frequency* | *Near unity* | *Transitional interaction* |
| *Hexagonal (Fig–A6)* | *Weak low-frequency* | *Smooth, low amplitude* | *Minimal* | *≈1 (flat)* | *Insufficient forcing* |
| *Vertical (harp) (Fig–A9)* | *Intermittent, discrete peaks* | *Alternating dominance* | *Sporadic* | *Sharp localized peaks* | *Incoherent interaction; poor capture* |

C. Physical interpretation

The spectral analysis establishes that droplet capture is governed not by the magnitude of fluctuations alone, but by their distribution in frequency and space. Efficient geometries are those that maintain coherent and moderately distributed spectral energy across the near-mesh region, enabling continuous droplet-flow interaction. In this context, the triangular mesh emerges as optimal because it achieves a balanced redistribution of fluctuation energy, ensuring that droplets experience sustained forcing without being subjected to excessive acceleration. Conversely, geometries that either concentrate energy too strongly (square) or fail to sustain it (hexagonal, vertical) exhibit reduced performance. These findings provide a mechanistic basis for the trends observed in Section 3.2 and form the basis for the dynamic matching argument developed in Section 3.4.

## 3.4 Dynamic matching between droplet response and flow structures

The spectral analysis in Section 3.3 demonstrates that capture efficiency depends on how mesh geometry redistributes fluctuation energy across frequency and space. In particular, the triangular mesh achieves optimal performance by sustaining broadband, moderately amplified fluctuations in both pore and obstruction regions (Fig. 8 and Table 4). To generalize this observation, the present section relates these spectral characteristics to the intrinsic response time of droplets.

The droplet response time, which characterizes the timescale over which a droplet adjusts to changes in the carrier-phase velocity, may be estimated (for small droplets in the Stokes regime) as

$$\tau_p = \frac{\rho_d D_d^2}{18\mu} \quad (12)$$

where $\rho_d$ is the droplet density, $D_d$ is the droplet diameter, and $\mu$ is the dynamic viscosity of air. This timescale governs the ability of droplets to deviate from fluid streamlines and respond to unsteady flow structures.

From the spectral analysis (Section 3.3), the flow field is characterized by a weighted mean frequency band rather than by a single universal frequency. A representative flow timescale is therefore defined as

$$\tau_f \sim \frac{1}{f_d} \quad (13)$$

where $f_d$ is the characteristic frequency extracted from the power spectral density (PSD). In the present study, $f_d$ is evaluated using a PSD-weighted averaging procedure that accounts for the entire spectral energy distribution, providing a representative flow timescale for both broadband and narrowband spectra. The detailed extraction methodology is provided in Appendix A4. This definition links the frequency-domain description of the flow directly to droplet dynamics.

The ratio of these two timescales defines a dynamic response parameter,

$$\Pi = \frac{\tau_p}{\tau_f} = \tau_p f_d \quad (14)$$

which quantifies the degree to which droplets can respond to geometry-induced unsteadiness. This parameter extends the conventional Stokes number interpretation by incorporating the actual spectral content of the flow, rather than relying solely on geometric length and velocity scales. It is important to note that since $f_d$ is extracted from the simulated flow field, the parameter Π is data-informed rather than fully a priori predictive. The key distinction from the classical Stokes number is that St uses a single geometric scale $l_f$ and mean velocity $U$, whereas Π uses the actual dominant oscillatory timescale of the near-mesh flow, which differs systematically between geometries even at the same *Re* and *U*. Geometries with broadband spectral content (e.g., the triangular mesh) and those with narrowband content (square) can share similar *St* values, yet exhibiting markedly different Π distributions, exemplifying why *St* alone cannot distinguish their capture efficiency trends. Although Π and the conventional Stokes number both contain the droplet response time, they differ in the definition of the flow timescale. The Stokes number compares droplet inertia with a characteristic timescale derived from the mean flow velocity and characteristic mesh length. In the present study, the inlet velocity, strand width ($l_f$ = 0.2 mm), and shade coefficient are identical across all geometries; consequently, *St* does not distinguish between the different mesh topologies for a given droplet size. In contrast, Π incorporates the characteristic frequency extracted from the PSD of the actual near-mesh flow field. As different mesh geometries generate different spectral signatures and flow timescales, Π varies between geometries even when *St* remains unchanged. Thus, Π should be interpreted as a frequency-based extension of the Stokes-number framework that incorporates geometry-dependent flow dynamics.

A. Regimes of droplet-flow interaction

The numerical results, interpreted through Figs. 6–7 and Table 4, indicate three distinct regimes: i) $\Pi \ll 1$(fast droplet response) - droplets rapidly adapt to flow fluctuations and remain strongly coupled to the carrier phase. As observed for small droplets in Fig. 6a (Zone I–II) and Fig. 6b (Regime I), this results in streamline-following behavior and low capture efficiency. ii) $\Pi \gg 1$(slow droplet response) - droplets are unable to respond to the dominant unsteadiness and follow nearly ballistic trajectories. In geometries with strong localized acceleration (e.g., square mesh; Table 4), this leads to rapid pore traversal and increased bypass, particularly in Zones II–III of Fig. 6a and in Regime II and IV of Fig. 6b. iii) $\Pi \sim O(1)$ - droplet response time becomes comparable to the flow timescale. In this regime, droplets partially respond to unsteady structures while retaining sufficient inertia to deviate from streamlines. This results in enhanced residence time within the near-mesh region (Zone III in Fig. 6a, Regime III in Fig 6b) and maximizes the probability of interception. In the present simulations, this regime is generally associated with $\Pi$ values of order unity (approximately 1–6), rather than a single critical value.

B. Connection to spectral characteristics and geometry

The role of geometry, as established in Section 3.3, is to shape the spectral distribution of the flow, thereby controlling the effective timescale $\tau_f$. The present results show that optimal capture is achieved when geometry produces a spectral environment that promotes the parameter $\Pi \sim O(1)$.

The triangular mesh satisfies this condition by generating broadband, moderately amplified fluctuations (Fig. 7), ensuring that a significant portion of the droplet population experiences dynamically compatible forcing. This leads to sustained interaction across the near-mesh region and explains its superior performance in Fig. 6. The square mesh, although characterized by strong spectral amplification, concentrates energy within narrow frequency bands (Table 4). This reduces the range of droplets satisfying the matching condition and promotes localized acceleration, thereby increasing bypass despite high instantaneous forcing. The trapezoidal mesh exhibits weaker and more uniformly distributed spectral content, resulting in partial matching but insufficient persistence of interaction. Consequently, capture efficiency remains intermediate. At the other extreme, the hexagonal mesh fails to generate sufficient spectral energy, leading to $\Pi \ll 1$for most droplets and weak inertial deviation. The vertical configuration, on the other hand, produces intermittent and incoherent spectral peaks, resulting in highly localized and temporally inconsistent forcing that does not sustain droplet–flow interaction.

C. Unified interpretation with Fig. 7

These observations provide a direct physical interpretation of the capture efficiency trends in Fig. 6. The zones identified in Fig. 6 correspond to regions where the interplay between droplet inertia and flow unsteadiness evolves spatially: In Zone I (upstream), droplets remain coupled to the flow ($\Pi \ll 1$). In Zone II (acceleration/shear layer), geometry-induced fluctuations begin to act on the droplets. In Zone III (near-mesh interaction), $\Pi \sim O(1)$ determines the extent of interception. In Zone IV (wake), residual unsteadiness influences post-interaction dynamics but does not significantly alter capture. Efficient geometries are therefore those that extend the spatial region over which $\Pi \sim O(1)$, thereby maximizing the duration and effectiveness of droplet–flow interaction.

D. Key implication

The analysis establishes that droplet capture is governed not only by inertia (Section 3.2) or spectral energy (Section 3.3), but by their dynamic compatibility. Geometry acts as a mediator by redistributing spectral energy in a manner that either promotes or inhibits this matching condition. This framework provides a unified explanation for the observed hierarchy of geometries and forms the basis for the reduced-order model developed in Section 3.5.

E. Sensitivity of Π to inlet velocity

To assess the sensitivity of the proposed framework to inlet velocity, additional Π values are evaluated at inlet velocities of 1, 3, and 5 m/s for representative droplet diameters corresponding to the four dynamic-response regimes identified in Figure 6(b). The results are summarized in Table 5. Although the characteristic frequency and corresponding Π values vary with both geometry and velocity, the representative droplet sizes remain within the same dynamic-response regimes across the investigated operating conditions. Small droplets (2 μm) consistently remain within the flow-following regime, intermediate droplets (15 μm) remain within the transitional regime, droplets associated with the highest capture efficiencies (30 μm) remain within the dynamic-matching regime, and larger droplets (40 μm) remain within the inertia-dominated regime. These observations suggest that the physical interpretation of the proposed framework is preserved over the investigated velocity range. Nevertheless, the present analysis should be regarded as a limited sensitivity assessment, and broader validation over additional velocities and turbulent inflow conditions remains an important subject for future work.

**TABLE 5.** Dynamic matching parameter (Π) evaluated for representative droplet diameters corresponding to the four response regimes identified in Fig. 6b at inlet velocities of 1, 3, and 5 m/s. The results demonstrate that although the exact Π values vary with mesh geometry and inlet velocity, the representative droplet sizes remain within the same dynamic-response regimes, supporting the robustness of the proposed spectral-dynamic framework over the investigated operating conditions.

| Velocity (m/s) | Droplet Size(μm) | Π value of different Fog mesh geometry type | | | | |
|---|---|---|---|---|---|---|
| | | Triangular | Trapezoidal | Hexagonal | Square | Vertical |
| 1 | 2 | 0.0146 | 0.0111 | 0.0076 | 0.0114 | 0.0049 |
| | 15 | 0.796 | 0.423 | 0.395 | 0.757 | 0.332 |
| | 30 | 3.22 | 2.06 | 1.36 | 2.87 | 1.15 |
| | 40 | 5.89 | 3.47 | 3.03 | 5.13 | 2.04 |
| 3 | 2 | 0.0151 | 0.0125 | 0.0083 | 0.0134 | 0.0048 |
| | 15 | 0.843 | 0.507 | 0.469 | 0.801 | 0.307 |
| | 30 | 3.49 | 2.33 | 1.88 | 3.14 | 1.35 |
| | 40 | 6.02 | 4.76 | 3.34 | 5.67 | 2.10 |
| 5 | 2 | 0.0153 | 0.0113 | 0.0107 | 0.0146 | 0.0054 |
| | 15 | 0.862 | 0.639 | 0.604 | 0.821 | 0.306 |
| | 30 | 3.45 | 2.56 | 2.42 | 3.29 | 1.22 |
| | 40 | 6.13 | 4.55 | 4.30 | 5.85 | 2.18 |

## 3.5 Physics-based correlation for capture efficiency

The results presented in Sections 3.2–3.4 demonstrate that droplet capture efficiency is governed by the dynamic compatibility between droplet inertia and geometry-induced flow unsteadiness, rather than by inertia alone. In particular, the spectral analysis (Section 3.3) and the zone-wise interpretation (Fig. 7) show that maximum capture occurs when the droplet response time becomes comparable to the characteristic timescale of the dominant flow structures, i.e., when $\Pi = \frac{\tau_p}{\tau_f} \sim O(1)$. To represent this behavior in a compact and predictive form, a physics-based correlation for capture efficiency is proposed as

$$\eta = \eta_{\max} \frac{\Pi}{\Pi + \Pi^{-1}} \exp\left[-\alpha\left(\frac{ln\Pi}{\beta}\right)^2\right], \tag{15}$$

where $\eta_{\max}$is the maximum attainable efficiency for a given geometry, and $\alpha$ is a parameter that characterizes the sensitivity of the system to mismatch between droplet and flow timescales.

The proposed form reflects the underlying physics of the three interaction regimes identified in Section 3.4, The term $\frac{\Pi}{\Pi+\Pi^{-1}}$ captures the monotonic increase in capture efficiency as droplets transition from a flow-following regime ($\Pi \ll 1$, Zone I–II in Fig.6) to an inertia-dominated regime. The exponential term $\exp\left[-\alpha(\frac{\ln \Pi}{\beta})^2\right]$ introduces a penalty for deviation from the optimal matching condition ($\Pi \sim O(1)$, Zone III), thereby accounting for the observed reduction in efficiency at large inertia (Zone IV), where bypass becomes dominant.

The logarithmic form is adopted because Π spans multiple decades and dynamic mismatch is more naturally interpreted on a multiplicative scale. Consequently, departures above and below the timescale-alignment condition ($\Pi \approx 1$) are treated in a symmetric manner in logarithmic space. The Gaussian penalty therefore provides a compact representation of the progressive reduction in capture efficiency as the droplet and flow timescales become increasingly mismatched. Unlike correlations based solely on the conventional Stokes number, the present formulation explicitly incorporates the flow timescale obtained from spectral analysis (Section 3.3). The correlation therefore accounts for geometry-dependent redistribution of fluctuation energy and the effective forcing experienced by droplets. The parameters $\eta_{max}$, $\alpha$ and $\beta$ are geometry-dependent and reflect each mesh's ability to sustain coherent, spatially persistant fluctuations (Table 6). The fitted values are summarized in Table 6. Geometries such as the triangular mesh, which exhibit broadband spectral content and balanced energy distribution, yield higher $\eta_{max}$ and lower sensitivity to mismatch. In contrast, geometries with localized or intermittent spectral characteristics (e.g., square or vertical configurations) exhibit stronger penalties when Π deviates from the droplet–flow compatibility. Accordingly Eq.(15) should be interpreted as a physics-inspired semi-empirical model rather than a first-principles derivation; additional data points, multi-velocity simulations, and experimental validation are needed before general predictive use.

**TABLE 6. Fitted parameters of the proposed efficiency correlation (Eq. 15) for different fog mesh geometries. Parameters are obtained by nonlinear least-squares fitting using MATLAB. The coefficient of determination ($R^2$) and 95% CI is included for each fit.**

| Geometry | $\eta_{max}$ (%)(95% CI) | $\alpha$(95% CI) | $\beta$(95% CI) | $R^2$ |
|---|---|---|---|---|
| Triangular | 34.1(32.28–36.25) | 1.35(1.11–1.58) | 0.58(0.42–0.66) | 0.979 |
| Square | 30.2(28.55–32.23) | 1.55(1.36–1.67) | 0.52(0.45–0.66) | 0.981 |
| Trapezoidal | 24.6(22.73–25.91) | 1.28(1.08–1.43) | 0.56(0.37–0.71) | 0.964 |
| Hexagonal | 15.8(13.09–17.99) | 1.10(0.79–1.36) | 0.63(0.41–0.79) | 0.936 |
| Vertical (Harp) | 9.6(7.01–11.76) | 1.70(1.21–2.1) | 0.50(0.32–0.78) | 0.942 |

## 3.6 Design implications for fog-harvesting meshes

The unified framework developed in Sections 3.2-3.5 provides physical design guidelines for efficient fog-harvesting meshes. The present five-geometry data set does not establish a universal optimum; rather, it indicates that high capture efficiency is favored when mesh geometry promotes timescale compatibility between droplets and flow structures between droplet response time and flow timescale over an extended near-mesh region (Zone III in Fig. 6). This requirement suggests the following design principles:

A. Spectral distribution over peak amplification

Efficient geometries should generate broadband, moderately amplified fluctuations rather than narrow-band or highly localized peaks. As shown in Section 3.3, the triangular mesh achieves superior performance by distributing spectral energy across both pore and obstruction regions, thereby sustaining droplet–flow interaction over a wide range of frequencies. In contrast, geometries that produce strong but localized amplification (e.g., square mesh) can induce excessive acceleration through the pores, increasing the likelihood of droplet bypass despite high instantaneous forcing.

B. Spatial distribution of unsteadiness

The results highlight the importance of maintaining spatially distributed flow structures across the near-mesh region. Distributed unsteadiness increases the effective residence time of droplets in Zone III (Fig. 6), enhancing the probability of interception. Geometries that concentrate fluctuations in either the pore region or the obstruction region alone are less effective, as they limit the spatial extent of droplet–flow interaction.

C. Avoidance of weak or intermittent forcing

Geometries that fail to generate sufficient spectral energy (hexagonal) or produce intermittent, incoherent fluctuations (vertical/harp configuration) exhibit poor performance. In such cases, droplets either remain fully entrained ($\Pi \ll 1$) or experience short-lived forcing that does not sustain interaction. This indicates that persistent, coherent unsteadiness is essential for effective capture.

D. Coupling between geometry and operating conditions

An important implication of the present framework is that favorable geometry is not defined solely by static pore shape, but also by its interaction with operating conditions. Since the flow timescale $\tau_f$ depends on the dominant frequency of the geometry-induced flow, and $f_d$ can scale with $U/l_f$ through Strouhal-type relations, Π is inherently flow-dependent. Consequently, the same mesh may perform differently under different wind speeds or droplet size distributions. Assuming approximate Strouhal-number similarity, the characteristic flow frequency is expected to scale as $f_d \propto U/l_f$. Consequently, Π will also vary with inlet velocity and characteristic mesh length. While this scaling suggests that the physical interpretation of Π should remain applicable across operating conditions, validation using multiple velocities is required before quantitative generalization can be established. Robust design therefore requires matching geometry to the expected operating envelope so that a significant fraction of the droplet population satisfies Π ~ O(1).

(v) Unified design criterion

The above considerations can be summarized into a single design objective: Maximize the spatial and spectral extent over which $\Pi \sim O(1)$is satisfied**.** This criterion provides a direct link between geometry, flow physics, and droplet dynamics, and offers a rational basis for optimizing fog-harvesting systems.

# 4. Conclusions

This study establishes a physics-based framework linking mesh geometry, spectral characteristics of the flow, and droplet capture efficiency in fog-harvesting systems. By combining Eulerian-Lagrangian simulations with frequency-domain analysis, the results demonstrate that capture efficiency is governed by the dynamic interaction between droplet inertia and geometry-induced unsteadiness, rather than by inertia or fluctuation intensity alone. The conclusions should be interpreted as design guidance from the five geometries examined, not as proof of a universal optimum. A limitation of the present study is the use of a uniform inlet velocity profile, which was adopted to isolate geometry-induced flow structures from externally imposed turbulence. Under realistic atmospheric conditions, inflow turbulence may modify the characteristic flow frequency and broaden the range of droplets satisfying the condition $\Pi \sim O(1)$. Furthermore, because the present simulations are performed at uniform inlet velocity, the proposed Π criterion should presently be interpreted as a mechanistic framework whose broader applicability requires validation across turbulent inflow conditions. Therefore, further validation under turbulent inflow conditions is recommended before extending the framework to field-scale predictions.

The key findings are summarized as follows:

1. Role of geometry in spectral redistribution: Mesh geometry fundamentally alters the distribution of fluctuation energy across frequency and space. Efficient geometries generate broadband, spatially distributed unsteadiness, whereas inefficient geometries either produce weak disturbances or highly localized, intermittent fluctuations.
2. Timescale compatibility criterion for enhanced capture: A characteristic flow timescale obtained from spectral analysis enables definition of a dynamic response parameter, $\Pi = \tau_p/\tau_f = \tau_p f_d$. Capture is enhanced when Π is order of unity, indicating timescale compatibility between droplet response and flow structures, indicating effective droplet–flow coupling. Unlike the conventional Stokes number, which remaines insensitive to mesh topology under the present conditions because the inlet velocity, characteristic strand width, and shade coefficient are held constant across all geometries, Π incorporates geometry-dependent spectral characteristics of the flow and therefore provides additional discrimination between different mesh configurations.

3. Zone-wise interpretation of droplet capture: The capture process can be interpreted in terms of distinct flow regions, with optimal interaction occurring in the near-mesh zone where droplets experience sustained forcing. Efficient geometries extend this region, thereby enhancing residence time and interception probability.
4. Unified correlation for capture efficiency: A physics-inspired semi-empirical correlation incorporating the dynamic matching parameter captures the observed dependence of efficiency on droplet inertia across the simulated geometries. The correlation is useful for interpretation and design guidance, while predictive use requires independent estimation of $f_d$ and validation across broader geometries and flow conditions.
5. Design implications: The spectral-dynamic framework provides physical design guidelines for fog-harvesting mesh geometries: improved capture performance tends to be associated with geometries that maximize the spatial and spectral extent over which the condition $\Pi \sim O(1)$ is satisfied, requiring moderate, broadband amplification of fluctuations rather than extreme or localized forcing. We emphasize that these are design guidelines derived from five discrete configurations; identification of a globally optimal geometry across the full design space would require parametric optimization using the Π criterion. Future work should extend the framework to: (i) multiple inlet velocities to validate the Strouhal-numbe r scaling of Π, (ii) turbulent inflow conditions representative of atmospheric fog environments, (iii) mesh arrays with varied porosity and strand topology, and (iv) probabilistic adhesion models incorporating rebound and splashing effects.

The present work shows that fog-harvesting efficiency is controlled by a spectral–dynamic coupling mechanism, in which geometry determines the flow timescale and droplets respond according to their intrinsic inertia. This perspective provides a unified and predictive framework for multiphase flow interactions with porous structures and offers a rational basis for the design of next-generation passive water-harvesting systems.

## ACKNOWLEDGEMENT

We acknowledge the National Supercomputing Mission (NSM) for providing computing resources of 'PARAM Sanganak' at IIT Kanpur, which is implemented by C-DAC and supported by the Ministry of Electronics and Information Technology (MeitY) and Department of Science and Technology (DST), Government of India. Also, we would like to thank the computer center (www.iitk.ac.in/cc ) at IIT Kanpur for providing the resources to carry out this work.

## AUTHOR DECLARATIONS

## CONFLICT OF INTEREST

The authors have no conflicts to disclose.

## DATA AVAILABILITY

The data that support the findings of this study are available from the corresponding author upon reasonable request.

# Appendix

### Appendix A1 - Coupling Algorithm

The Eulerian and Lagrangian components of the solver are progressed sequentially within each global time step via a robustly coupled two-way interaction strategy executed utilizing the PIMPLE algorithm. At the commencement of each time step, the carrier-phase flow field is revised utilizing the interphase momentum source terms derived from the preceding iteration. Thereafter, the scattered droplets are advanced by Lagrangian particle tracking with sub-timestepping to guarantee numerical stability and precise trajectory integration. At this stage, the forces applied by the droplets on the adjacent fluid are assessed and aggregated. The reaction forces are subsequently assigned to the relevant Eulerian control volumes to revise the momentum source term. To reduce temporal lag between the two phases, one or more outer PIMPLE corrector iterations are executed, facilitating the convergence of the Eulerian and Lagrangian solutions inside the same global time step. This repeated coupling approach guarantees continuous momentum transfer between the two phases and enhances the stability of the coupled solution[12]. A flowchart(Fig A1) illustrating the coupling technique is included to enhance comprehension of the overall computing sequence.

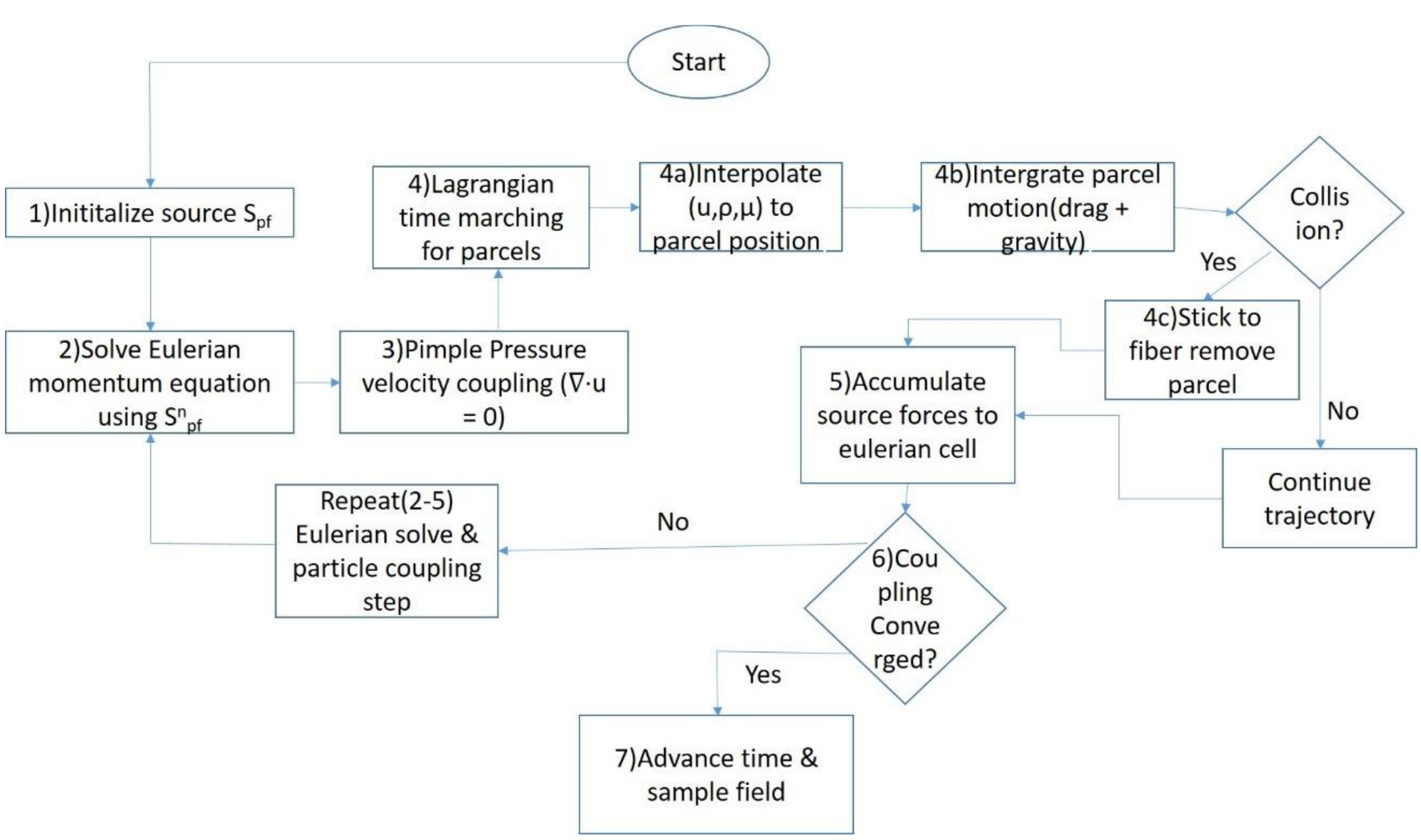

Fig. A1: A schematic representation of the coupling procedure is provided in the accompanying flowchart to facilitate understanding of the overall computational sequence.

**Appendix A2 - Fog Droplet distribution**

Fog droplets generally possess sizes varying from around 2 μm to 40 μm, as indicated by field observations of natural fog occurrences[22,40]. The diverse range of droplet sizes generates a vast array of particle inertial reactions in relation to fiber dimensions and incoming wind velocity, significantly affecting interception, impaction, and aerodynamic capture on the fog mesh. The diversity in the inlet droplet population is represented by employing a limited two-parameter Rosin–Rammler (RR) distribution for the droplet size distribution.

$$F(d) = 1 - \exp\left[-\left(\frac{d}{d_\eta}\right)^n\right], \qquad d \in [\, d_{min}\,, d_{max}] \tag{16}$$

where *F(d)* represents the cumulative fraction of droplets with diameter smaller than $d$, $d_\eta$ denotes the characteristic scale parameter, and n controls the spread of the distribution. The bounds are set to $d_{min}$ = 2μm and $d_{max}$ = 40μm to remain consistent with experimentally observed fog droplet spectra.

The Rosin–Rammler distribution is extensively employed to characterize liquid droplet ensembles, as it offers analytical formulations for distribution moments, facilitating the constant alignment of macroscopic parameters such as number concentration, liquid water content, and effective droplet diameter. Moreover, its linearized form facilitates accurate parameter estimation from discretized droplet observations[6,38]. Parameter calibration is additionally informed by recent in-situ measurements from the SIRTA atmospheric observatory[32], which indicate a unimodal fog droplet size distribution with peak concentrations generally found between 11 μm and 25 μm. The parameters $d_\eta$ and $n$ in Eq. 16 are modified within the limited RR framework to replicate a peak within the experimentally reported range, as depicted in Fig. A2.

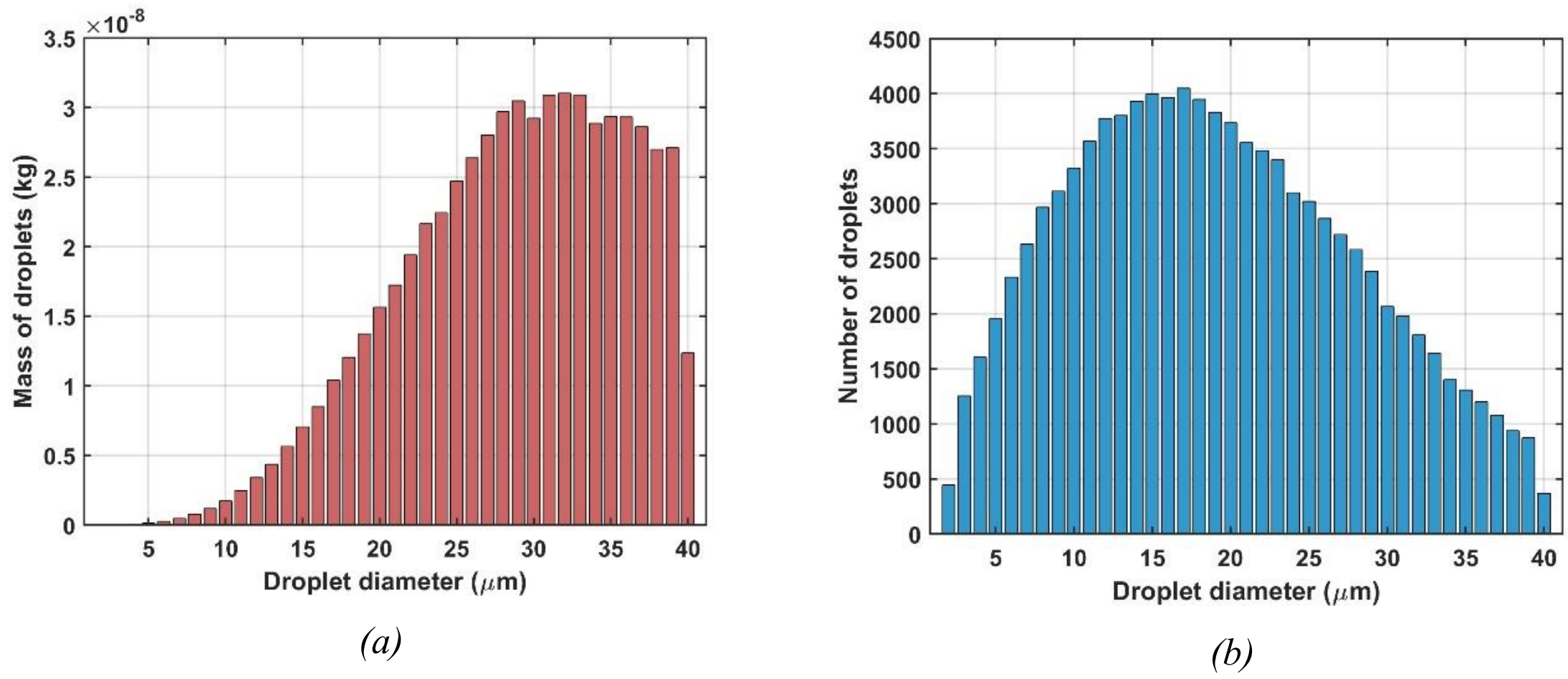

FIG A2. Rosin Rammler distribution of fog droplets used in our study a) Rosin Rammler mass distribution of the droplets, b) Percentage of droplets shown in numbers with peaks between 10 to 25μm

Figures A2(a) and A2(b) illustrate the resultant droplet population utilized in the simulations, represented by number distribution and mass distribution, respectively. Although smaller droplets prevail in quantity, larger droplets account for the majority of the liquid water mass because droplet mass is cubically dependent on diameter. This distinction is crucial for fog harvesting models, as droplet inertia—and hence capture probability—is closely associated with droplet size. The resultant parameterization offers a concise and experimentally coherent depiction of the inlet fog droplet spectrum, while maintaining direct compatibility with the Rosin–Rammler injection model utilized in OpenFOAM, rendering it highly appropriate for the current Eulerian–Lagrangian simulations of droplet–mesh interaction.

**Appendix A3 - Modeling the capture efficiency**

In actual meteorological conditions, the interaction between fog and a collector transpires through multiple sequential stages. As fog droplets near the collecting surface, the airflow is disrupted by the mesh, producing aerodynamic forces that redirect some of the arriving droplets. Droplets with smaller inertia adhere to the streamlines that go around the mesh fibers and hence pass through the mesh holes without engaging with the fibers. On the other hand, droplet with high inertia diverge significantly from the airflow, resulting in collisions with the mesh structure and subsequent capture. Upon contact with the mesh fibers, droplets merge with previously deposited droplets and progressively aggregate, presumably forming a very thin film on the fiber. As the liquid volume increases, the total droplet mass ultimately attains a critical threshold at which gravity forces surpass surface adhesion, permitting the accumulated water to flow downward into the collection channel[14].

Due to the multi-scale nature of fog–mesh interactions, fog harvesting performance is commonly described using three efficiency components—namely aerodynamic, deposition, and drainage efficiencies—as introduced earlier. Rather than redefining these processes, we adopt the same framework here for quantitative evaluation. The overall collection efficiency is expressed as:

$$\eta_{tot} = \eta_a \times \eta_{de} \times \eta_{dr} \quad (17)$$

This formulation assumes independence among the three mechanisms, although minor coupling may arise in practice due to factors such as pore blockage, mesh deformation, and local flow redistribution. Nevertheless, this multiplicative framework remains widely accepted for isolating dominant physical contributions to fog collection.

In the present study, the OpenFOAM solver directly evaluates the capture efficiency, defined as the combined effect of aerodynamic and deposition efficiencies,

$$\eta_{cap} = \eta_a \times \eta_{de} \quad (18)$$

$\eta_{cap}$ is ascertained by introducing Lagrangian droplets at the inlet under undisturbed upstream flow conditions and monitoring their paths during interaction with the mesh. A droplet that contacts a fiber is presumed to adhere deterministically, implying that rebound is excluded from the current model. Under this premise, droplet interaction with a fiber is regarded as analogous to deposition. Thus, $\eta_{cap}$ is determined as the ratio of the droplets that adhere to the mesh to the droplets injected across the upstream frontal region of the collector, whose undisturbed trajectories would meet the mesh. The total collection efficiency can be calculated as $\eta_{tot} = \eta_{cap} \times \eta_{dr}$ , with $\eta_{dr}$ assessed independently. The current study does not explicitly simulate the drainage processes, as modeling post-deposition liquid films, droplet re-entrainment under shear forces, and incomplete dripping at collector edges introduces additional multiscale physical phenomena involving surface science and fluid dynamics, entailing uncertain parameters that necessitate focused investigation.

**Appendix A4 - Frequency-Domain Analysis**

The unstable aerodynamic behavior caused by droplet-flow interaction around the fog mesh is examined by analyzing velocity fluctuations in the carrier phase by Fast Fourier Transform (FFT) and Power Spectral Density (PSD) techniques in the frequency domain. Spectral analysis is extensively employed in fluid dynamics to discern predominant oscillation modes and delineate unstable flow structures in turbulent or vortex-dominated flows[39,45]. In droplet-laden flows, the interaction between scattered droplets and the carrier airflow can create localized velocity variations that affect droplet trajectories and capture behavior. The spectrum analysis facilitates the comparison between the carrier-phase flow and the droplet-laden flow field, enabling the identification of frequency ranges where droplet-flow coupling alters the distribution of kinetic energy in the velocity fluctuations. Frequency-domain analysis thus offers more understanding of the temporal attributes of the flow structures forming around the fog mesh. Eighteen probes are systematically set in multiple groups along the principal flow direction to document the evolution of flow disturbances caused by the interaction between the mesh and droplets. These probe groups delineate certain flow regions in relation to the mesh, encompassing areas far upstream of the mesh, the upstream acceleration zone, directly upstream of the mesh surface, immediately downstream of the mesh, the downstream wake region, and areas far downstream of the collector. Numerous probes were strategically positioned within each region at slightly varied lateral locations to capture localized fluctuations in the velocity field, while concentrating on the core area of the mesh where droplet-flow interaction is most significant. The velocity signals acquired from these probe locations serve as the foundation for the ensuing spectral analysis.

The instantaneous velocity data are recorded during the statistically stationary phase of the simulation, following the dissipation of initial transients. The overall sampling length was $T = 0.06$ s, during which $N = 12{,}000$ samples were recorded for each probe. This corresponds to a sample interval of $\Delta t = 5 \times 10^{-6}$ s. The sampling window T = 0.06 s is adequate for capturing fiber-scale vortex shedding: at U = 5 m/s and $l_f \approx 0.2$ mm, the Strouhal-number estimate gives $f \approx 5000$ Hz, yielding ∼300 shedding cycles within T. However, domain-scale low-frequency interactions (f < 100 Hz) are less well resolved within this window and the corresponding spectral content should be interpreted with caution. The streamwise velocity component was obtained at each probe position and utilized for spectral analysis, as changes in the streamwise direction are most strongly linked to the acceleration and deceleration of the flow interacting with the fog mesh. Before conducting spectral analysis, the mean velocity component was eliminated from the data to isolate the variable velocity component. If $u(t)$ signifies the instantaneous velocity signal, the fluctuating component is defined as $u'(t) = u(t) - \bar{u}$ , where $\bar{u}$ represents the time-averaged velocity at the probe position. Eliminating the mean component guarantees that the spectral analysis exclusively reflects the dynamic variations of the flow field[7]. The discrete Fourier transform of the variable signal is subsequently calculated to turn the time-domain signal into its frequency-domain equivalent. The Fourier transform of a discrete signal of N samples can be articulated as

$$\hat{u}(f_k) = \sum_{n=0}^{N-1} u'(t_n) e^{-i2\pi kn/N} \tag{19}$$

Where $u'(t_n)$ represents the fluctuating velocity at time $t_n$, $k$ denotes the frequency index, and the corresponding discrete frequency given by $f_k = \frac{k}{N\Delta t}$. To obtain a statistically robust estimate of the spectral energy distribution, the power spectral density is computed using Welch's method[47]. In this approach, the velocity signal is divided into overlapping segments of equal length. Each segment is multiplied by a window function and its individual periodogram is computed using FFT. The final PSD is obtained by averaging the spectra from all segments, thereby reducing the variance associated with single-FFT estimates.

In the present study, each signal segment contains $L$ = 4096 samples with a 50% overlap between consecutive segments. A Hann window was applied to each segment to reduce spectral leakage associated with finite-length sampling. The Hann window is defined as

$$w(n) = \frac{1}{2}\left(1 - cos\left(\frac{2\pi n}{L-1}\right)\right) \tag{20}$$

The windowed velocity signal is therefore expressed as $u_w(n) = w(n)u'(n)$. The Fourier transform of each windowed segment yields the segment-specific spectrum

$$\hat{u}_m(f) = \sum_{n=0}^{L-1} u_m(n)e^{-i2\pi fn/L} \tag{21}$$

For which the corresponding segment power spectrum is computed as

$$PSD_m(f) = \frac{|\hat{u}_m(f)|^2}{L} \tag{22}$$

The final Welch-averaged power spectral density is then obtained by averaging the spectra of all $M$ segments,

$$PSD(f) = \frac{1}{M}\sum_{m=1}^{M} PSD_m(f) \tag{23}$$

window-energy normalization was applied to ensure that the spectral amplitude remains physically consistent after windowing.

The resulting Power Spectral Density (PSD) represents the distribution of kinetic energy associated with velocity changes throughout the frequency spectrum. According to the aforementioned sampling parameters, the spectrum analysis yields a frequency resolution of $\Delta f = \frac{f_s}{N} = 16.7\ Hz,$ with the maximum resolvable frequency constrained by the Nyquist frequency $f_{max} = \frac{f_s}{2}$. The exchange of momentum of droplets on the carrier flow is quantified by analyzing spectral values obtained from the power spectral density (PSD) of velocity fluctuations. The PSD of the droplet-laden flow, referred to as *S1*, signifies the spectral energy distribution of velocity variations in the presence of droplets, while *S2* pertains to the PSD of the single-phase flow devoid of droplets. The spectral gain is defined as the ratio of droplet-laden to single-phase spectral energies, with *S1/S2* representing the two-way coupling spectral gain designated as $G_1$. This number quantifies the enhancement of velocity variations resulting from droplet-flow interaction. Values of *S1/S2* > 1 signify that the presence of droplets enhances the spectral energy of the flow at a certain frequency, whereas values near unity suggest minimal droplet impact. In addition to spectral gain, the droplet-induced energy difference is evaluated using the difference between the two spectra,

$$\Delta PSD = S1 - S2 \tag{24}$$

where *ΔPSD* represents the change in spectral energy due to droplet injection. Positive values of *ΔPSD* indicate that droplets introduce additional kinetic energy into the flow at that frequency, whereas negative values indicate a reduction in spectral energy. Together, the spectral gain and *ΔPSD* provide complementary information regarding the strength of droplet–flow coupling and the redistribution of turbulent energy across different frequency bands. All spectral processing, including signal conditioning, FFT computation, and PSD estimation, is performed using custom post-processing scripts implemented in MATLAB (MathWorks Inc.). The probe velocity data are first extracted from the OpenFOAM simulation using the built-in probe sampling utility and subsequently processed in MATLAB to obtain the frequency-domain representations of the flow signals. The characteristic flow frequency, $f_d$, is obtained from the PSD-weighted mean frequency (spectral centroid),

$$f_d = \frac{\sum_i f_i S_i}{\sum_i S_i} \tag{25}$$

where $f_i$ is the frequency and $S_i$ is the corresponding PSD amplitude. The summation is performed over the analyzed frequency range of 100–5000 Hz. This range excludes poorly resolved low-frequency content while encompassing the mesh-scale flow structures identified from the Reynolds-number and Strouhal-number analysis. The PSD-weighted mean frequency was selected because several mesh geometries exhibit broadband spectral distributions rather than a single dominant peak. By incorporating contributions from the entire spectral energy distribution, the spectral centroid provides a consistent measure of the characteristic flow timescale across all geometries. The resulting $f_d$ values are subsequently used in the calculation of $\Pi$ is described in Section 3.4.

Appendix A5 - Frequency-Domain Characterization of Droplet–Flow Interaction of hexagonal mesh

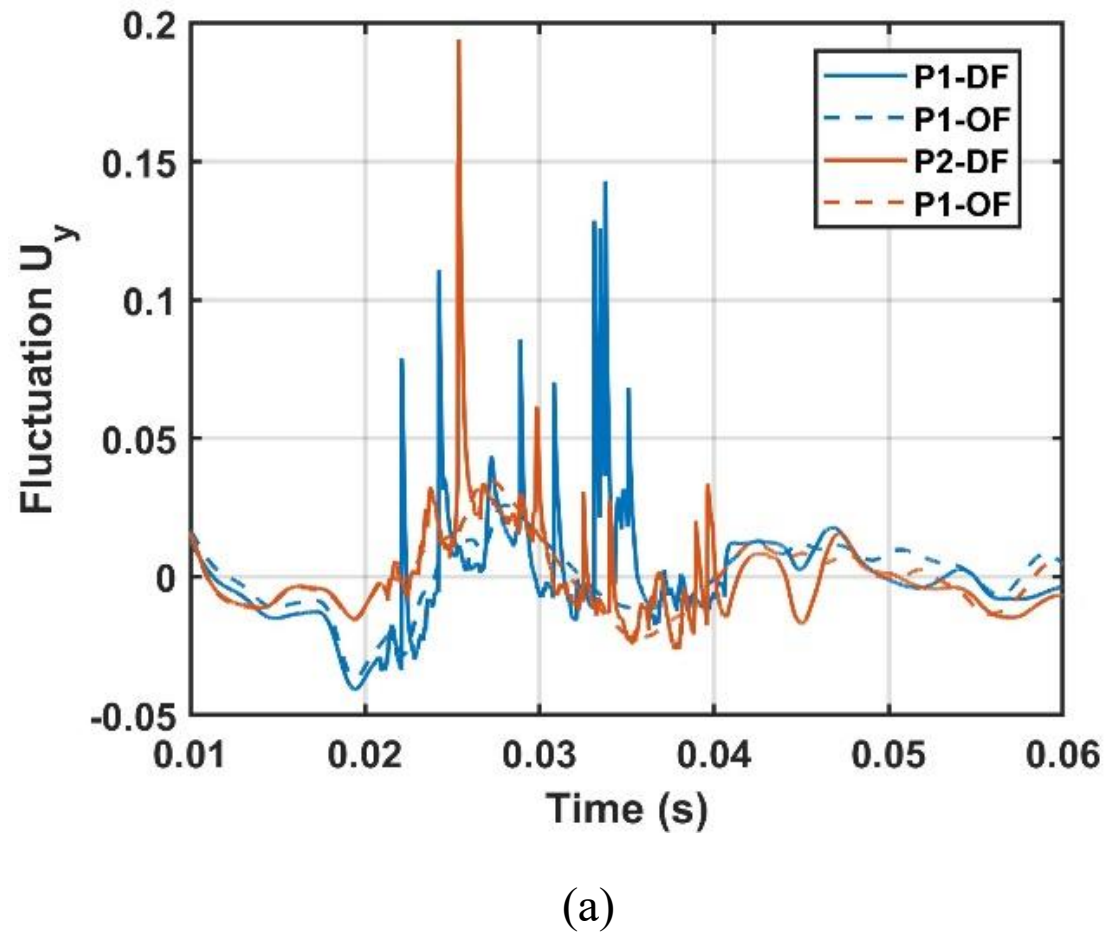

(a)

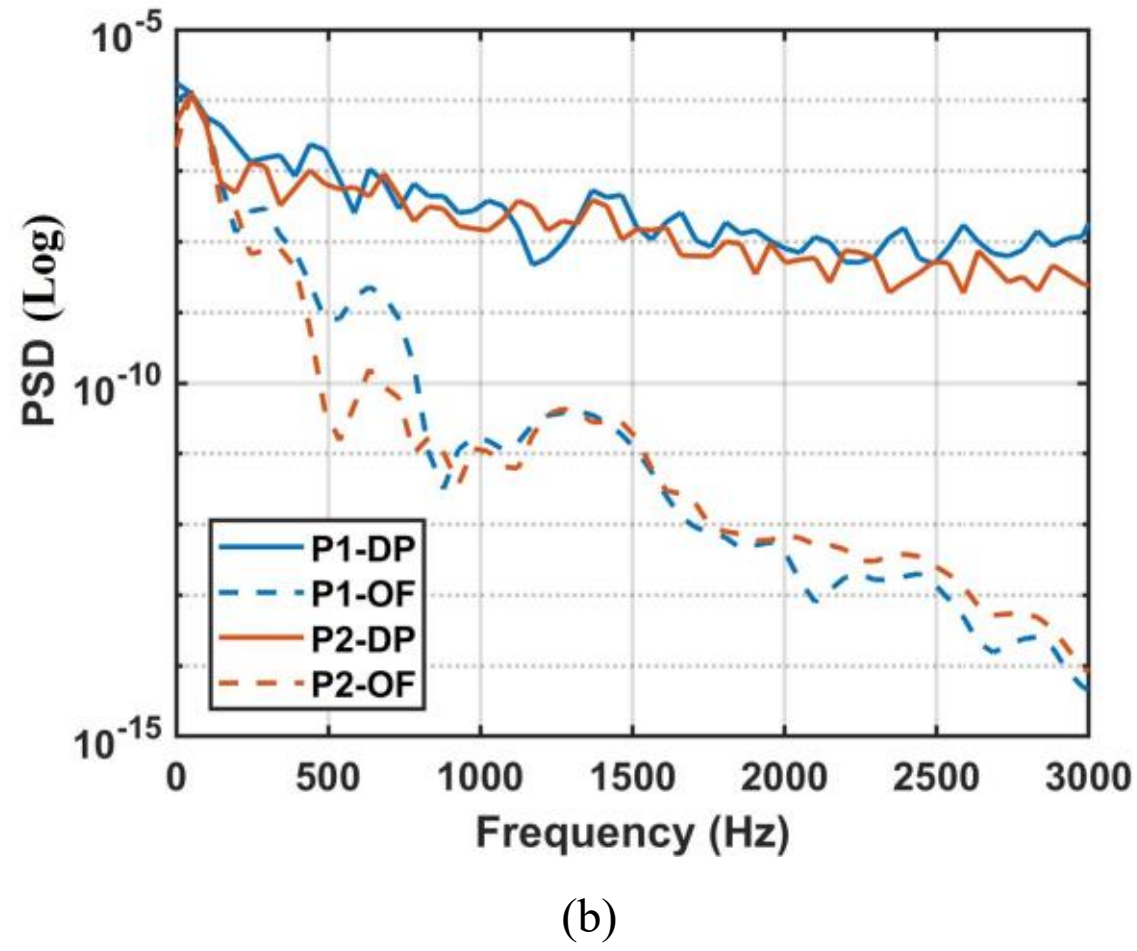

(b)

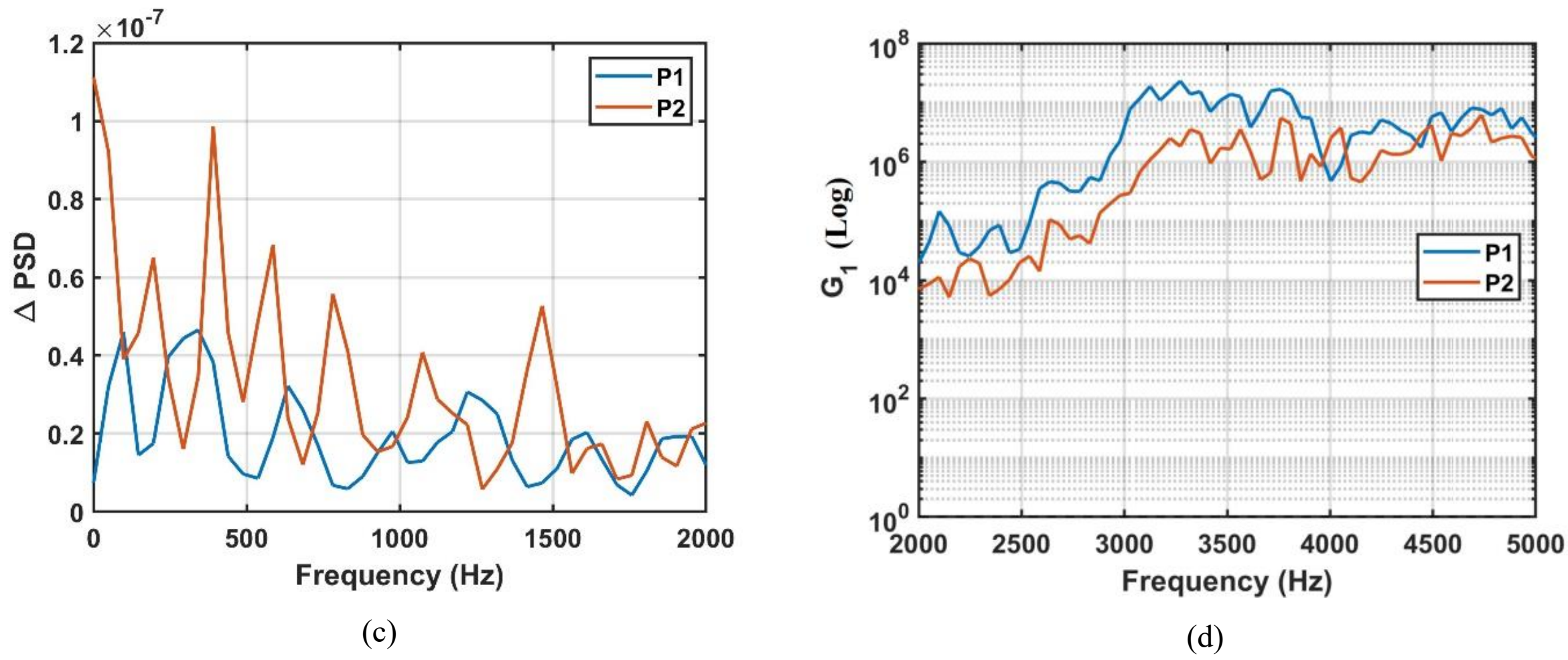

Fig. A5 Frequency-domain characterization of droplet–flow interaction for the hexagonal mesh: (a) Fourier spectrum of transverse velocity fluctuations, (b) power spectral density (PSD) comparing droplet-laden (DF) and droplet-free (OF) cases, (c) droplet-induced spectral variation ΔPSD, and (d) spectral gain $G_1 = S_1/S_2$. Results are shown for probes located in the obstruction (P1) and pore (P2) regions.

Appendix A6 - Frequency-Domain Characterization of Droplet–Flow Interaction of trapezoidal mesh

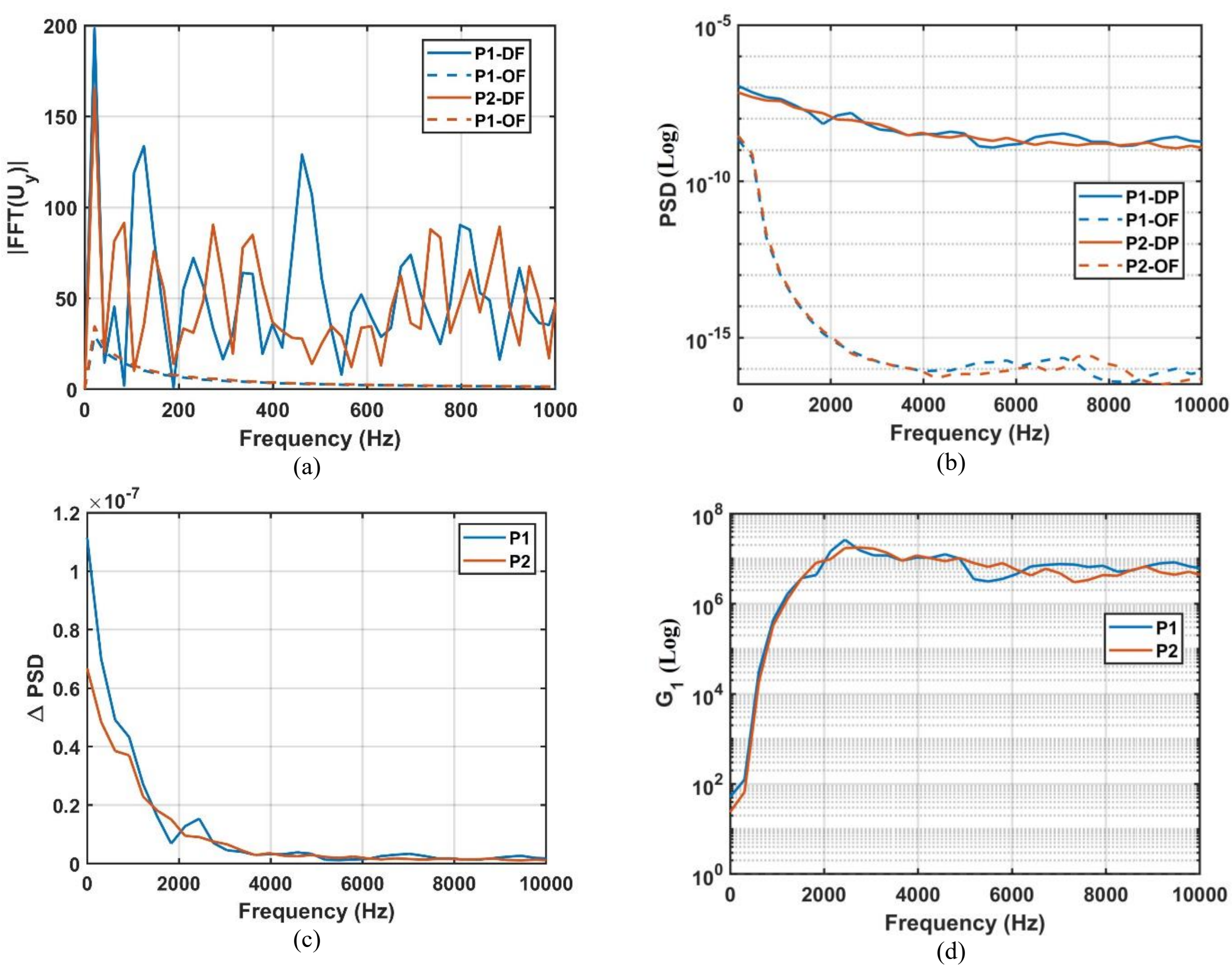

Fig. A6 Frequency-domain characterization of droplet–flow interaction for the trapezoidal mesh: (a) Fourier spectrum of transverse velocity fluctuations, (b) power spectral density (PSD) comparing droplet-laden (DF) and droplet-free (OF) cases, (c) droplet-induced spectral variation ΔPSD, and (d) spectral gain $G_1 = S_1/S_2$. Results are shown for probes located in the obstruction (P1) and pore (P2) regions.

Appendix A7 - Frequency-Domain Characterization of Droplet–Flow Interaction of square mesh

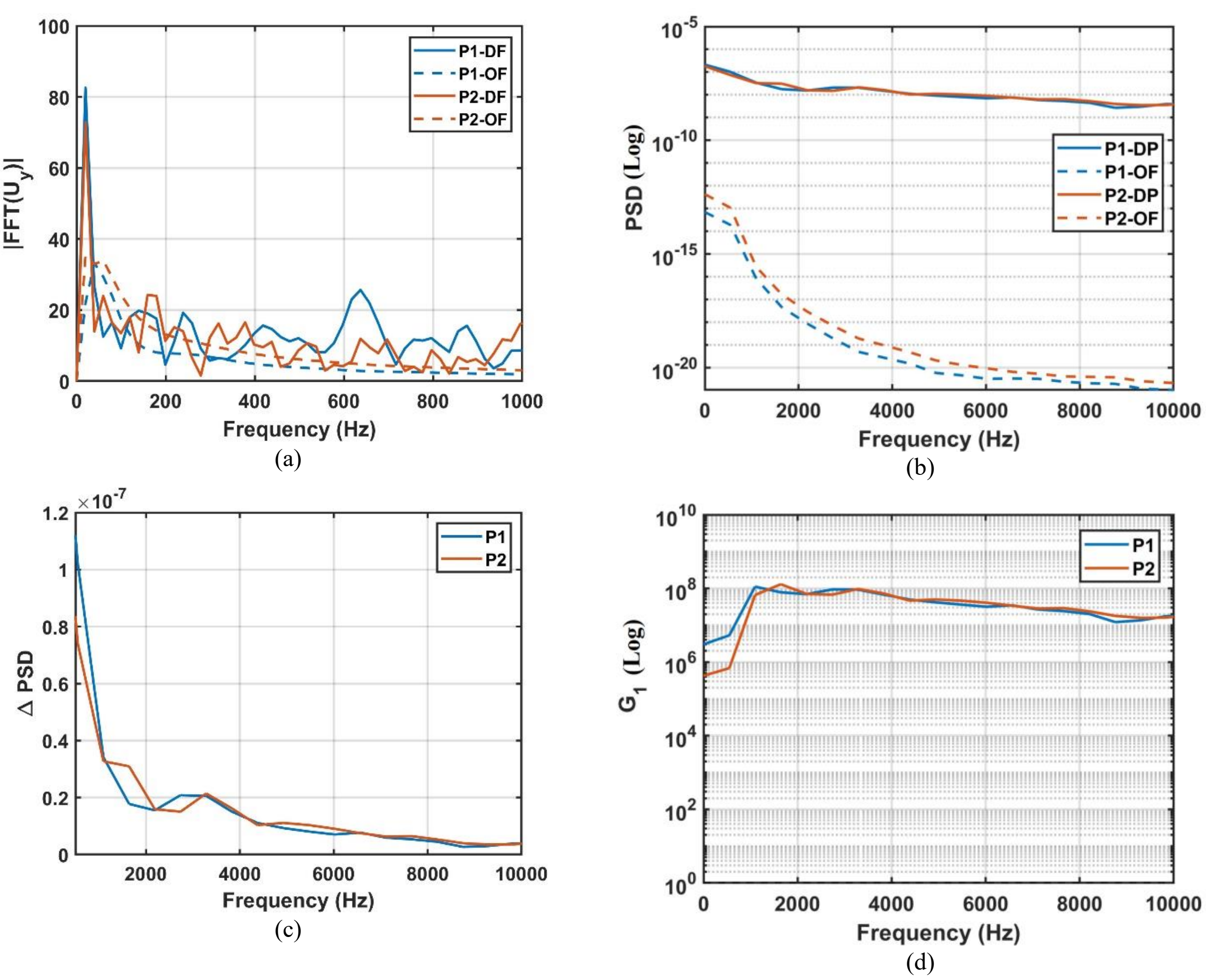

Fig. A7 Frequency-domain characterization of droplet–flow interaction for the square mesh: (a) Fourier spectrum of transverse velocity fluctuations, (b) power spectral density (PSD) comparing droplet-laden (DF) and droplet-free (OF) cases, (c) droplet-induced spectral variation ΔPSD, and (d) spectral gain $G_1 = S_1/S_2$. Results are shown for probes located in the obstruction (P1) and pore (P2) regions.

Appendix A8 - Frequency-Domain Characterization of Droplet–Flow Interaction of vertical mesh

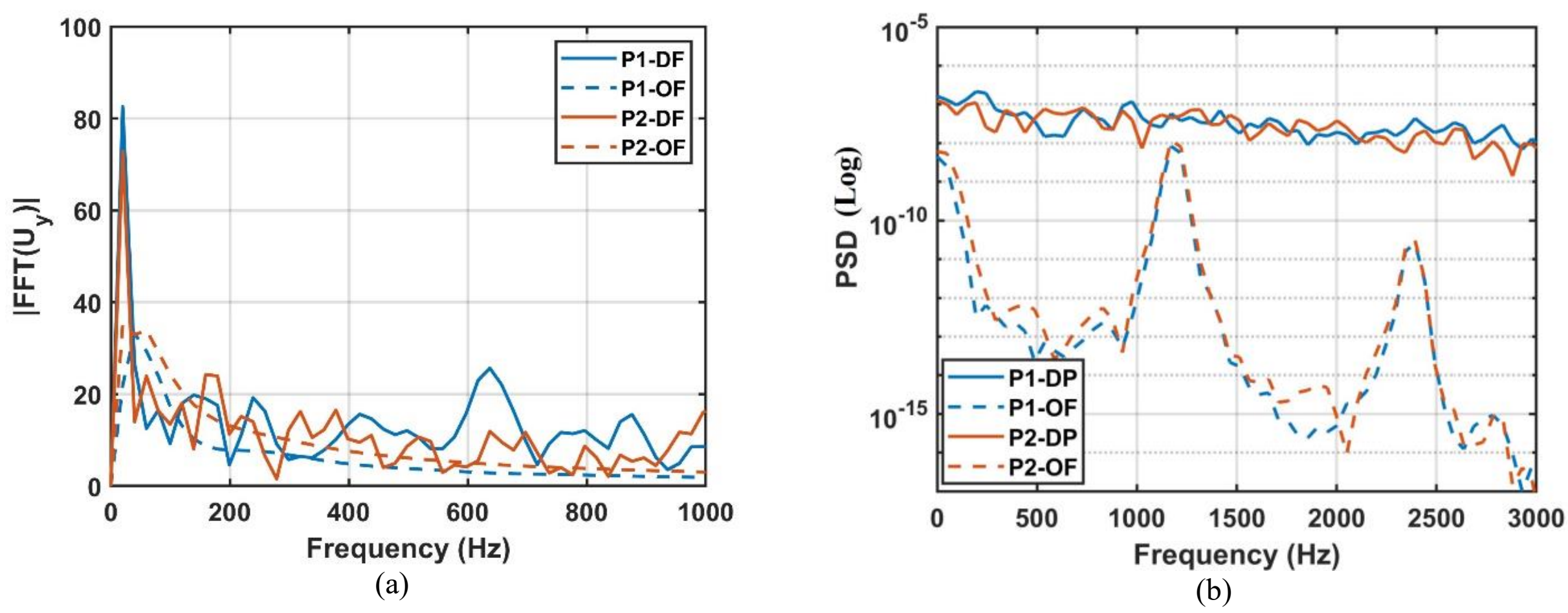

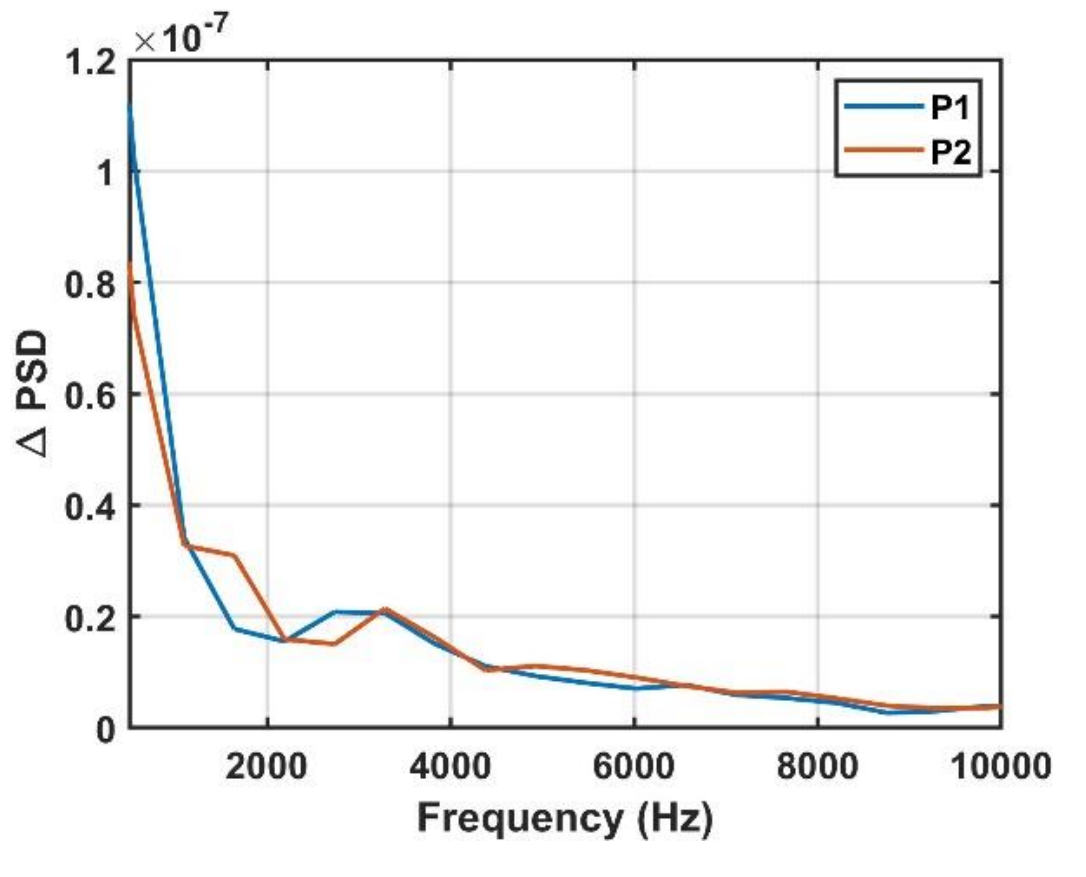

(c)

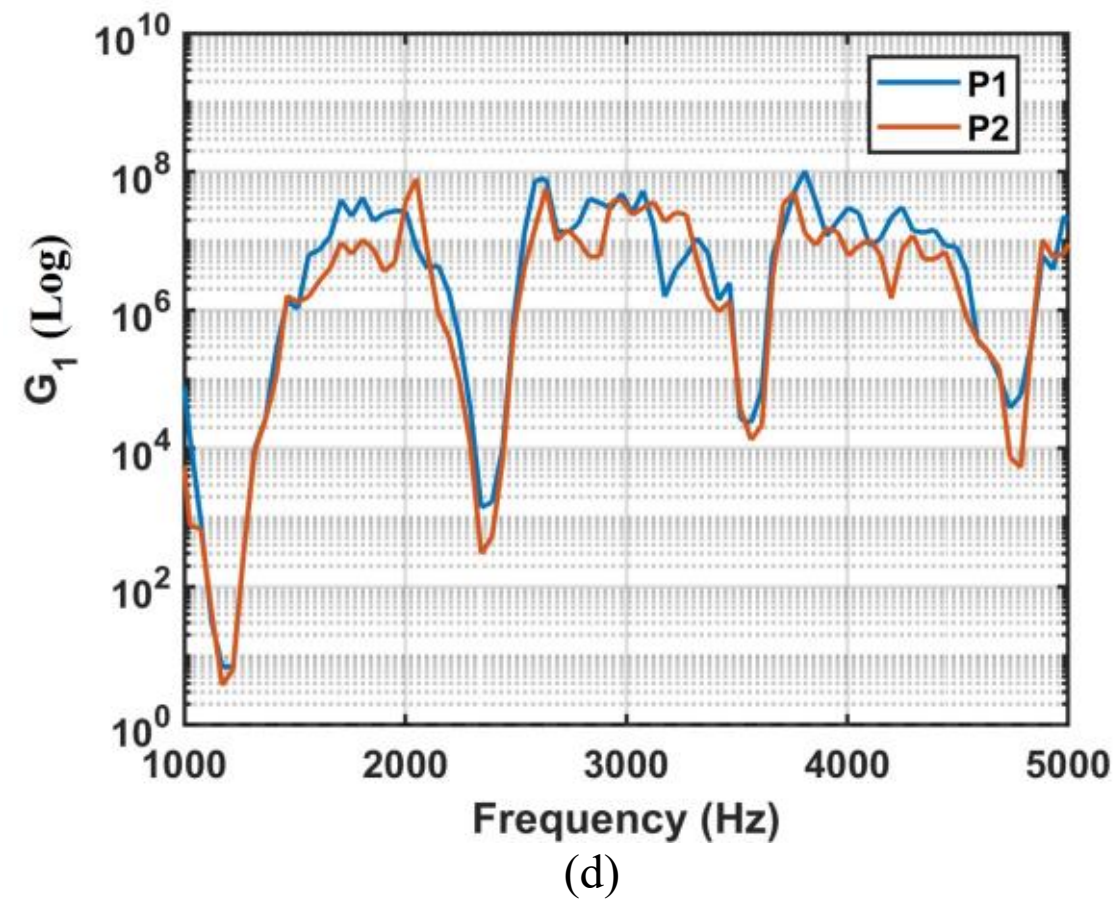

(d)

Fig. A8 Frequency-domain characterization of droplet–flow interaction for the vertical mesh: (a) Fourier spectrum of transverse velocity fluctuations, (b) power spectral density (PSD) comparing droplet-laden (DF) and droplet-free (OF) cases, (c) droplet-induced spectral variation ΔPSD, and (d) spectral gain $G_1 = S_1/S_2$. Results are shown for probes located in the obstruction (P1) and pore (P2) regions.